\documentclass[prb,twocolumn,showpacs,preprintnumbers,amsmath,aps]{revtex4-1}
\usepackage{graphicx,hyperref}% Include figure files
\usepackage{xcolor}
\usepackage{bm}% bold math
\usepackage{subfigure}% bold math
\usepackage{amsthm}
\def\nbC{{\mathchoice {\setbox0=\hbox{$\displaystyle\rm C$}%
\hbox{\hbox to0pt{\kern0.4\wd0\vrule height0.9\ht0\hss}\box0}}
{\setbox0=\hbox{$\textstyle\rm C$}\hbox{\hbox
to0pt{\kern0.4\wd0\vrule height0.9\ht0\hss}\box0}}
{\setbox0=\hbox{$\scriptstyle\rm C$}\hbox{\hbox
to0pt{\kern0.4\wd0\vrule height0.9\ht0\hss}\box0}}
{\setbox0=\hbox{$\scriptscriptstyle\rm C$}\hbox{\hbox
to0pt{\kern0.4\wd0\vrule height0.9\ht0\hss}\box0}}}}
\def\nbQ{{\mathchoice {\setbox0=\hbox{$\displaystyle\rm
Q$}\hbox{\raise
0.15\ht0\hbox to0pt{\kern0.4\wd0\vrule height0.8\ht0\hss}\box0}}
{\setbox0=\hbox{$\textstyle\rm Q$}\hbox{\raise
0.15\ht0\hbox to0pt{\kern0.4\wd0\vrule height0.8\ht0\hss}\box0}}
{\setbox0=\hbox{$\scriptstyle\rm Q$}\hbox{\raise
0.15\ht0\hbox to0pt{\kern0.4\wd0\vrule height0.7\ht0\hss}\box0}}
{\setbox0=\hbox{$\scriptscriptstyle\rm Q$}\hbox{\raise
0.15\ht0\hbox to0pt{\kern0.4\wd0\vrule height0.7\ht0\hss}\box0}}}}
\def\nbT{{\mathchoice {\setbox0=\hbox{$\displaystyle\rm
T$}\hbox{\hbox to0pt{\kern0.3\wd0\vrule height0.9\ht0\hss}\box0}}
{\setbox0=\hbox{$\textstyle\rm T$}\hbox{\hbox
to0pt{\kern0.3\wd0\vrule height0.9\ht0\hss}\box0}}
{\setbox0=\hbox{$\scriptstyle\rm T$}\hbox{\hbox
to0pt{\kern0.3\wd0\vrule height0.9\ht0\hss}\box0}}
{\setbox0=\hbox{$\scriptscriptstyle\rm T$}\hbox{\hbox
to0pt{\kern0.3\wd0\vrule height0.9\ht0\hss}\box0}}}}
\def\nbS{{\mathchoice
{\setbox0=\hbox{$\displaystyle     \rm S$}\hbox{\raise0.5\ht0%
\hbox to0pt{\kern0.35\wd0\vrule height0.45\ht0\hss}\hbox
to0pt{\kern0.55\wd0\vrule height0.5\ht0\hss}\box0}}
{\setbox0=\hbox{$\textstyle        \rm S$}\hbox{\raise0.5\ht0%
\hbox to0pt{\kern0.35\wd0\vrule height0.45\ht0\hss}\hbox
to0pt{\kern0.55\wd0\vrule height0.5\ht0\hss}\box0}}
{\setbox0=\hbox{$\scriptstyle      \rm S$}\hbox{\raise0.5\ht0%
\hboxto0pt{\kern0.35\wd0\vrule height0.45\ht0\hss}\raise0.05\ht0%
\hbox to0pt{\kern0.5\wd0\vrule height0.45\ht0\hss}\box0}}
{\setbox0=\hbox{$\scriptscriptstyle\rm S$}\hbox{\raise0.5\ht0%
\hboxto0pt{\kern0.4\wd0\vrule height0.45\ht0\hss}\raise0.05\ht0%
\hbox to0pt{\kern0.55\wd0\vrule height0.45\ht0\hss}\box0}}}}
\def\nbZ{{\mathchoice {\hbox{$\sf\textstyle Z\kern-0.4em Z$}}
{\hbox{$\sf\textstyle Z\kern-0.4em Z$}}
{\hbox{$\sf\scriptstyle Z\kern-0.3em Z$}}
{\hbox{$\sf\scriptscriptstyle Z\kern-0.2em Z$}}}}

\makeatletter
\def\@fnsymbol#1{\ensuremath{\@alph{#1}}} % a, b, c … in the title
\makeatother

\begin{document}

\title{The effect of droplet configurations within the Functional Renormalization Group of the Ising model approaching the lower critical dimension}

\author{Ivan Balog} \email{balog@ifs.hr}
\affiliation{Institute of Physics, P.O.Box 304, Bijeni\v{c}ka cesta 46, HR-10001 Zagreb, Croatia}

\author{Lucija Nora Farka\v{s}} \email{lufarkas@irb.hr}
\affiliation{Institute of Physics, P.O.Box 304, Bijeni\v{c}ka cesta 46, HR-10001 Zagreb, Croatia}

\author{Maroje Marohni\'{c}}\email{maroje.marohnic@gmail.com}
\affiliation{Independent researcher, Zagreb, Croatia}

\author{Gilles Tarjus} \email{tarjus@lptmc.jussieu.fr}
\affiliation{LPTMC, CNRS-UMR 7600, Sorbonne Universit\'e, 4 Place Jussieu, 75252 Paris cedex 05, France}

\date{\today}

\begin{abstract}

We explore the application of the nonperturbative functional renormalization group (NPFRG), within its most common approximation scheme 
based on truncations of the derivative expansion, to the $Z_2$-symmetric scalar $\varphi^4$ theory as the lower critical dimension $d_{\rm lc}$ 
is approached. We aim to assess whether the NPFRG —a broad, nonspecialized method which is accurate in $d\geq 2$— can capture the effect 
of the localized (droplet) excitations that drive the disappearance of the phase transition in $d_{\rm lc}$ and control the critical behavior as 
$d\to d_{\rm lc}$. We extend a prior analysis to the next (second) order of the derivative expansion to check the convergence 
of the results and the robustness of the conclusions. The study turns out to be much more involved. 
Through extensive numerical and analytical work we provide evidence that the convergence to $d_{\rm lc}$ is nonuniform in the field dependence 
and is characterized by the emergence of a boundary layer near the minima of the fixed-point effective potential. This is the mathematical 
mechanism through which the NPFRG within the truncated derivative expansion reproduces nontrivial features predicted by the droplet theory 
of Bruce and Wallace\cite{bw81_lett,bw83}, namely, the existence of two distinct small parameters as $d\to d_{\rm lc}$ that control 
different aspects of the critical behavior and that are nonperturbatively related.  The second order of the derivative expansion 
fixes several issues that were encountered at the lower level and improves the compatibility with the droplet-theory predictions.

\end{abstract}

\pacs{11.10.Hi, 75.40.Cx}

\maketitle

\section{Introduction}
\label{sec:introduction}

We wish to understand how the nonperturbative functional renormalization group (NPFRG), which is a powerful and versatile approach to 
study the emergence of scale invariance in a variety of complex systems,\cite{berges02,dupuis21} fares at describing situations in which the 
long-distance physics is controlled by highly nonuniform field configurations and localized excitations. We focus on one of the simplest yet physically 
relevant examples, the approach to the lower critical dimension in the $Z_2$-symmetric scalar $\varphi^4$ theory which is in the same 
universality class as the Ising model. 

The lower critical dimension of the Ising model is exactly known to be $d_{\rm lc}=1$  from the Peierls argument.\cite{peierls36} The 
one-dimensional physics at low temperature is dominated by the proliferation of localized, instantonic, excitations in the form of kinks and 
antikinks which provide the mechanism for destroying the finite-temperature phase transition. The approach to the lower critical 
dimension has also been described in detail by Bruce and Wallace\cite{bw81_lett,bw83} through a ``droplet" theory, based on the observation 
that the low-dimensional behavior of the model is controlled by coarse-grained configurations that include droplets of one phase or the other 
on all scales. At criticality these droplets proliferate and have a nonzero density, their distribution being ruled by the (renormalized) surface 
tension. Close enough to the lower critical dimension $d_{\rm lc}=1$, the distribution of droplets can be considered in the dilute limit, which 
allows Bruce and Wallace to derive the full droplet distribution. (More precisely, what is then dilute is the droplet boundaries that become in 
exactly $d_{\rm lc}=1$ the kinks and antikinks corresponding to the one-dimensional instantons.) 

A key outcome of the droplet theory is that the critical physics in low dimensions is controlled by two distinct characteristics of the droplets, the 
fractal dimension of the droplet surface and the critical droplet concentration, which both go to zero  as one approaches the lower critical 
dimension $d_{\rm lc}=1$ but are nonperturbatively related. Denoting by $\epsilon=d-1$ the distance from the lower critical dimension, the 
former is simply given by $\epsilon$ while the latter goes as $\exp(-2/\epsilon)/\epsilon$. The presence of two small parameters, exponentially 
(or reciprocally logarithmically) related, is therefore a hallmark of the approach to the lower critical dimension of Ising-like models.

The question that we are interested in is whether the long-distance physics resulting from droplets, kinks and antikinks can be captured by 
the NPFRG which is a theoretical tool based on generic approximation schemes such as the so-called ``derivative expansion". The latter 
focuses on uniform coarse-grained configurations and slowly varying spatial fluctuations around them that are described through an expansion 
in gradients of the field. 

There have been previous attempts to study the Ising model or the $Z_2$-symmetric scalar $\varphi^4$ theory in $d=1$ through the NPFRG 
and the truncated derivative expansion, but they have only found partial success. Interest in the problem comes from the fact that it also 
describes the quantum barrier tunneling of a particle in a double well potential. The counterpart of the absence of finite-temperature transition 
is that a particle can never be localized in one minimum for an infinite time even if the barrier height is arbitrarily large. Several 
works\cite{aoki02,kapoyannis00} have investigated whether the NPFRG  could capture the dependence  on the barrier height $V$ of the energy 
gap between the first excited state and the ground state, a dependence that is known to go as $\exp(-1/V)$. However, whereas good agreement 
with exact results has been found for small barrier heights,\cite{aoki02} the predictions worsen as the barrier height increases, even 
when improving the approximation.\cite{weryauch06,zappala01,bonanno22} Arguments have been given that the truncated derivative expansion 
in the NPFRG actually cannot reproduce the asymptotic low-temperature scaling associated with the proliferation of kinks and antikinks.\cite{rulquin16} 
Also in $d=1$, attempts to optimize the infrared regulator present in the NPFRG formalism have been unable to recover the (exact) absence 
of symmetry breaking.\cite{nandori14} A similar problem appears in the NPFRG treatment of a resistively shunted Josephson junction for which 
one does not correctly capture all regimes of the Schmidt transition.\cite{daviet23}

The issue with the above one-dimensional studies is that nothing guarantees in the NPFRG approach involving approximations such as a 
truncated derivative expansion that the lower critical dimension is indeed equal to $1$. In models where the symmetry that is broken 
at the transition is continuous as in O($N>2$) field theories, the lower critical dimension $d_{\rm lc}=2$ and  the appearance of Goldstone modes 
is easily captured by the approximations of the NPFRG. In contrast, the value of  $d_{\rm lc}$ when the broken symmetry is discrete  
depends on the details of the approximation scheme. 

In a previous work we have studied the approach to the lower critical dimension and the value of $d_{\rm lc}$ in the $Z_2$-symmetric scalar 
$\varphi^4$ theory within the lowest relevant truncation of the derivative expansion which is known as LPA'\,\cite{lnf23}. At odds with an earlier 
NPFRG study,\cite{bbw04} we have shown that the convergence of the fixed-point effective action when the dimension $d\to d_{\rm lc}$ is 
nonuniform in its field dependence. (Note that this property can only be found because the RG in use is {\it functional}.) It is characterized 
by the emergence of a boundary layer around the minima of the fixed-point effective potential whose width goes to zero and whose location 
moves out to infinity. By means of a singular perturbation analysis we have characterized this behavior and derived an explicit analytical 
expression for the lower critical dimension $d_{\rm lc}$ whose value depends on the chosen IR regulator and can be fine-tuned to be 
$1$ by tweaking the latter. An important outcome of the boundary-layer mechanism is that the approach to $d_{\rm lc}$ is not described 
by a unique small parameter but instead involves two logarithmically related vanishing scales, much as predicted by the droplet theory of 
Bruce and Wallace.

In the present work we extend this analysis of the LPA' approximation to the second order of the derivative expansion. This is of course a 
necessary step to assess the robustness and generality of our results. Furthermore,  because of its simplicity, the LPA' 
approximation features some unsatisfactory properties. One of them is that the anomalous dimension of the field $\eta$, which is central for an 
analysis of the lower critical dimension, crucially depends on the choice of renormalization prescription, while this is no longer true at the 
second order of the derivative expansion.  The study, however, turns out to be considerably more difficult. We have gathered strong numerical 
evidence for the emergence of a boundary layer much like in the LPA' approximation but the singular perturbation treatment is much more 
involved and we could not manage full numerical and analytical control. As will be presented in detail below, we have nonetheless combined 
analytical arguments and numerical data to convincingly support the existence of a boundary-layer solution for the fixed-point functions and 
characterize its properties. The critical behavior obtained as $d\to d_{\rm lc}$ captures several key results of the droplet theory,\cite{bw81_lett,bw83}  
including the behavior of the correlation length exponent $\nu$ that could not be reproduced within the LPA' approximation.
This confirms that at least some of the long-distance physics controlled by strongly nonuniform coarse-grained configurations, involving here 
droplets on all scales, can be captured within the NPFRG by approximation schemes that at a first glance appear mostly blind to these configurations.

%, except for the already mentioned fact that $d_{\rm lc}$ is not generically equal to the exact value of $1$.

The outline of the paper is as follows. In Section \ref{sec:FRG} we give a summary of the NPFRG formalism applied to the approach 
toward the lower critical dimension of the $Z_2$-symmetric scalar $\varphi^4$ theory and we introduce the second-order of the derivative 
expansions, denoted by $\partial^2$. In Sec.~\ref{sec_nonuniform} we present the singular perturbation treatment of the coupled fixed-point 
equations for the two functions of the field present at the $\partial^2$ order and we give numerical evidence for a nonuniform convergence 
that is characterized by the emergence of a boundary layer around the minima of the effective potential as $d\to d_{\rm lc}$. We next 
analyze in detail in Sec.~\ref{sec:BL} the solution of the equations in the boundary layer. In Sec.~\ref{sec:matching_de2} we consider the 
difficult problem of finding the solution in the region where the field is of O(1) and of matching it with the solution in the boundary layer. 
Based on the solution in the LPA' approximation and some further analytical results we put forward a scenario that appears corroborated by 
numerical data and we give an estimate of $d_{\rm lc}$ for several choices of infrared regulator. The stability of the fixed point and the behavior 
of the eigenvalues when $d\to d_{\rm lc}$ is considered in Sec.~\ref{sec:essential}. Most of the discussion focuses on whether the inverse of the 
correlation length exponent $1/\nu$ goes to zero as $d\to d_{\rm lc}$. This is indeed a key property of the exact solution and a prerequisite for 
observing ``essential scaling" in the lower critical dimension. Finally in Sec \ref{sec:conclusion} we summarize the results and give some 
concluding remarks. The main text is complemented by several appendices where we provide additional detail.
\\

\section{Functional RG approach}
\label{sec:FRG}

The Ising universality class is represented in a field-theoretical setting by the $Z_2$-symmetric scalar $\varphi^4$ action
\begin{equation}
\label{eq_bare-action}
S[\varphi]=\int_x \big [ \frac{1}{2}(\partial_\mu \varphi(x))^2 + \frac{r}{2} \varphi(x)^2 + \frac{u}{4!}\varphi(x)^4 \big ],
\end{equation}
where $\int_x\equiv \int d^d x$. We study the critical behavior of this model through the NPFRG \cite{berges02,dupuis21} which is a modern 
incarnation of Wilson's renormalization group.\cite{wilson74} In this approach one uses an infrared (IR) regulator function, 
\begin{equation}
\label{eq_regulator}
\Delta S_k[\varphi]=\frac{1}{2} \int_{xy} R_k(x-y)\varphi(x)\varphi(y),
\end{equation}
with the function $R_k$ obeying several criteria discussed at length in previous works,\cite{litim01, de6,canet03} to suppress integration over 
the fluctuations with momentum less than some running cutoff $k$ in the functional integral defining the partition function of the theory at 
equilibrium. With the addition of this regulator the partition function now reads 
\begin{equation}
Z_k[J]=\int \mathcal D\varphi \exp[-S[\varphi]-\Delta S_k[\varphi] +\int_x J(x)\varphi(x)].
\end{equation}

The cutoff-dependent effective average action, i.e., the cutoff-dependent generating functional of the 1-particle irreducible (1PI) correlation 
functions, is given by a modified Legendre transform 
\begin{equation}
\label{eq_Legendre}
\Gamma_k[\phi]=-\ln Z_k[J]+\int_x J(x)\phi(x)-\Delta S_k[\phi],
\end{equation}
with $\phi(x)=\langle \varphi(x)\rangle=\delta \ln Z_k[J]/\delta J(x)$. The flow of the effective average action with the cutoff $k$ obeys an 
exact FRG equation\cite{wetterich93,wetterich93a,wetterich93b}
\begin{equation}
\label{eq_ERGE}
\partial_t\Gamma_k[\phi]=\frac 12\int_{xy}\partial_t R_k(x-y)\big [(\Gamma_k^{(2)}[\phi]+R_k)^{-1}\big ]_{xy},
\end{equation}
where $\Gamma_k^{(2)}$ is the second functional derivative of $\Gamma_k$ and $t=\ln(k/\Lambda)$ with $\Lambda$ a UV cutoff. At the 
UV scale, $\Gamma_k[\phi]$ essentially reduces to the bare action whereas when $k=0$ all the fluctuations have been integrated and one 
recovers the exact effective action (or Gibbs free-energy functional in the language of magnetic systems).

The exact FRG equation in Eq.~(\ref{eq_ERGE}) is a convenient starting point for nonperturbative approximations.\cite{bmw, canet03,de6} 
A commonly used and generic approximation scheme is based on the truncation of the derivative expansion, 
\begin{equation}
\label{eq_DE}
\Gamma_k[\phi]=\int_x \big [ U_k(\phi(x))+\frac{1}{2}Z_k(\phi(x))(\partial_x \phi(x))^2 + \cdots \big ],
\end{equation}
where the ellipses denote terms involving a higher number of derivatives of the field. In the present work we focus on the second-order 
approximation (then dropping all the terms appearing in the ellipses), which we refer to as $\partial^2$ for short. Note that this approximation 
scheme attempts to capture the long-distance physics of the problem by expanding around uniform configurations of the average (also called 
``classical" or ``background") field.

After inserting the $\partial^2$ ansatz in the exact FRG equation, Eq.~(\ref{eq_ERGE}), one obtains two coupled nonlinear partial differential 
equations for the functions $U_k(\phi)$ and $Z_k(\phi)$, which are the cutoff-dependent effective potential and field renormalization function 
respectively. At criticality scale invariance is associated with a fixed point of the FRG flow which can be found by first introducing scaling 
dimensions and dimensionless quantities,
\begin{equation}
\phi=\frac{k^{(d-2)/2}}{Z_{k}^{1/2}}\varphi,\;\; U_k(\phi)=k^du_k(\varphi),\;\; Z_k(\phi)=Z_{k} z_k(\varphi),
\end{equation}
where the flow of $Z_{k}$ defines the running anomalous dimension, 
\begin{equation}
\eta_k=-\frac{\partial_t Z_{k}}{Z_{k}}.
\end{equation}
If a fixed point is indeed found, $\eta_k\to \eta_*$ when $k\to 0$, and the scaling dimension of the field is then given by 
\begin{equation}
\label{eq_field-dim}
D_\phi=\frac{(d-2+\eta_*)}{2}.
\end{equation} 
(Note that we have used the same notation $\varphi$ for the bare variable in Eq.~(\ref{eq_bare-action}) and the dimensionless average 
field, as the former will no longer appear in what follows.) 

When expressed in terms of dimensionless quantities, the FRG equations for the effective potential and the field renormalization function 
take the form
\begin{equation}
\begin{aligned}
\label{eq_ERGEdimensionless}
&\partial_t u_k(\varphi)=-d u_k(\varphi) +\frac{(d-2+\eta_k)}{2}\varphi u_k'(\varphi) + \beta^{(d)}_{u}(\varphi;\eta_k) \\&
\partial_t z_k(\varphi) = \eta_k z_k(\varphi) +\frac{(d-2+\eta_k)}{2}\varphi z_k'(\varphi) + \beta^{(d)}_{z}(\varphi;\eta_k),
\end{aligned}
\end{equation}
where a prime indicates a derivative with respect to the argument of the function and $\beta^{(d)}_u(\varphi)$ and $\beta^{(d)}_z(\varphi)$ are 
functions of $w_k(\varphi)=u_k''(\varphi)$, $z_k(\varphi)$, and of their first two derivatives; they are given in Appendix~\ref{appx_flows}. The 
running anomalous dimension $\eta_k$ is given by the renormalization condition, $z_k(\varphi=0)=1$, i.e.,
\begin{equation}
\label{eq_running_eta}
\eta_k=- \beta^{(d)}_{z}(\varphi=0;\eta_k).
\end{equation}
The fixed point is reached when $t=\ln(\frac{k}{\Lambda})\to -\infty$ (or $k\to 0$) and the left-hand sides of the equations in 
Eqs.~(\ref{eq_ERGEdimensionless}) go to zero while $\eta_k\to \eta_*$.

In the present study we have used several forms of regulator with the dimensionless IR cutoff function $r(y=q^2/k^2)=R_k(q^2)/(Z_k q^2)$ 
chosen as
\begin{equation}
\begin{aligned}
\label{eq_cutoff_choices}
&r_{\Theta}(y)=\alpha \Theta(1-y)(1-y)/y \\&
r_e(y)=\alpha e^{-y}/y \\&
r_W(y)=\alpha \frac{1}{e^{y}-1},
\end{aligned}
\end{equation}
where $\Theta$ is the Heaviside step function and $\alpha$ is a variational parameter of order 1.  We refer to the above choices as ``Theta" 
or ``Litim"\cite{litim01}, and ``Exponential" \cite{de6}, and ``Wetterich"\cite{wetterich93a} cutoff functions.

For a given value of the dimension $d$ and a chosen IR cutoff function, we can then numerically solve the fixed-point equations and obtain 
$u_*(\varphi)$, $z_*(\varphi)$, and $\eta_*$. For simplicity, we will drop the asterisk that characterizes fixed-point quantities in most of the following.

Here we are interested in the approach to the lower critical dimension $d_{\rm lc}$ within the $\partial^2$ approximation. 
We have already studied in a previous work\cite{lnf23} a lower order of the derivative expansion, the LPA',  in which one describes the 
effective action through the effective potential $U_k(\phi)$ and a scale-dependent but field-independent field renormalization function $Z_{k}$ 
defined through the renormalization condition that $Z_{k}=Z_k(\phi_{{\rm min},k})$ with $\phi_{{\rm min},k}$ the location of the minimum 
of the effective potential. (Note that due to the $Z_2$ symmetry one can restrict the analysis to the positive fields, the situation for negative 
fields being obtained by symmetry.)

Finally, we comment on a minor point concerning our terminology. When considering exact RG flow equations as Eq.~(\ref{eq_ERGE}) we speak 
of exact RG (ERG) or functional RG (FRG). After having introduced nonperturbative {\it approximations} in the exact FRG, and provided we still 
deal with functions of the field, we rather use nonperturbative FRG (NPFRG). 
\\

\section{Nonuniform convergence to the lower critical dimension}
\label{sec_nonuniform}

\subsection{Fixed-point equations as $d$ approaches $d_{\rm lc}$}

When discussing the fixed point of the FRG flow for $d$ approaching the lower critical dimension $d_{\rm lc}$, it is convenient to introduce the 
small parameter $\tilde{\epsilon}$,
\begin{equation}
\label{eq_tildeepsilon_def}
\tilde\epsilon(d)=\frac{d-2+\eta(d)}{2(2-\eta(d))}=\frac{D_\phi}{d-2 D_\phi},
\end{equation}
where the second equality follows from Eq.~(\ref{eq_field-dim}). One of the hallmarks of a lower critical dimension is that the field does not rescale, i.e., $D_\phi=0$. Its fluctuations always remain of order $1$ in the thermodynamic limit, which prevents ordering. The approach to the lower 
critical dimension is then described by $\tilde\epsilon \to 0$.

As one expects the relation between $d$ and $\tilde\epsilon$ to be monotonic, one can substitute $\tilde\epsilon$ for $d$ as a control parameter, 
which is what we do in our analytic investigations. The starting point of the latter is therefore the set of two coupled differential equations for 
$u(\varphi)$ and $z(\varphi)$, or more conveniently for the square mass function $w(\varphi)=u''(\varphi)$ and $z(\varphi)$,
\begin{equation}
\begin{aligned}
\label{eq_ERGEdimensionless_FP}
0=&- w(\varphi) +\tilde\epsilon\varphi w'(\varphi) + \left(\frac{1+2\tilde\epsilon}{d(\tilde\epsilon)}\right)\beta^{(d(\tilde\epsilon))}_{w}(\varphi;\eta(\tilde\epsilon)) \\
0 =& \eta(\tilde\epsilon) \left(\frac{1+2\tilde\epsilon}{d(\tilde\epsilon)}\right) z(\varphi) +\tilde\epsilon\varphi z'(\varphi)\\& + \left(\frac{1+2\tilde\epsilon}{d(\tilde\epsilon)}\right)\beta^{(d(\tilde\epsilon))}_{z}(\varphi;\eta(\tilde\epsilon))
\end{aligned}
\end{equation}
where $\beta^{(d)}_w(\varphi)=\partial^2_\varphi \beta^{(d)}_u(\varphi)$, $d(\tilde\epsilon)=(1+2\tilde\epsilon)(2-\eta(\tilde\epsilon))$. These two 
equations are complemented by an equation for the minimum of the effective potential $\varphi_{\rm min}$ obtained from $u'(\varphi_{\rm min})=0$ 
as
\begin{equation}
\label{eq_ERGEdimensionless_min_FP}
0 =\tilde\epsilon\varphi_{\rm min} w(\varphi_{\rm min}) +\left(\frac{1+2\tilde\epsilon}{d(\tilde\epsilon)}\right)
\partial_\varphi \beta^{(d(\tilde\epsilon))}_{u}(\varphi;\eta(\tilde\epsilon))\vert_{\varphi=\varphi_{\rm min}}
\end{equation}
and the renormalization prescription for $\eta$: $z(\varphi=0)=1$.

When $\tilde\epsilon\to 0$, $d\to d_{\rm lc}$ and $\eta\to 2-d_{\rm lc}$, with $d_{\rm lc}$ now being a quantity to be determined which, due to the 
approximation, may differ from the exact value $d_{\rm lc}=1$. One can see that the second term of the ordinary differential equations in 
Eqs.~(\ref{eq_ERGEdimensionless_FP}) {\it a priori} vanishes when $\tilde\epsilon\to 0$. However, one must be careful because the limit may not 
be uniform in the field $\varphi$.

This is precisely what we found in our study of the lower-order approximation LPA'.\cite{lnf23} We observed that the approach to the lower critical 
dimension involves a boundary layer in the effective potential $u(\varphi)$ around its minima, $\pm \varphi_{\rm min}$. (Actually, the layers are 
not quite at the boundary and should rather be called ``interior layers", but we keep the terminology ``boundary layers" which is more common.) 
Such problems can be handled through the singular perturbation theory,\cite{KevorkianCole1981, Hunter2004} which we discuss below.

\begin{figure}
    \centering
    \includegraphics[width=1\linewidth]{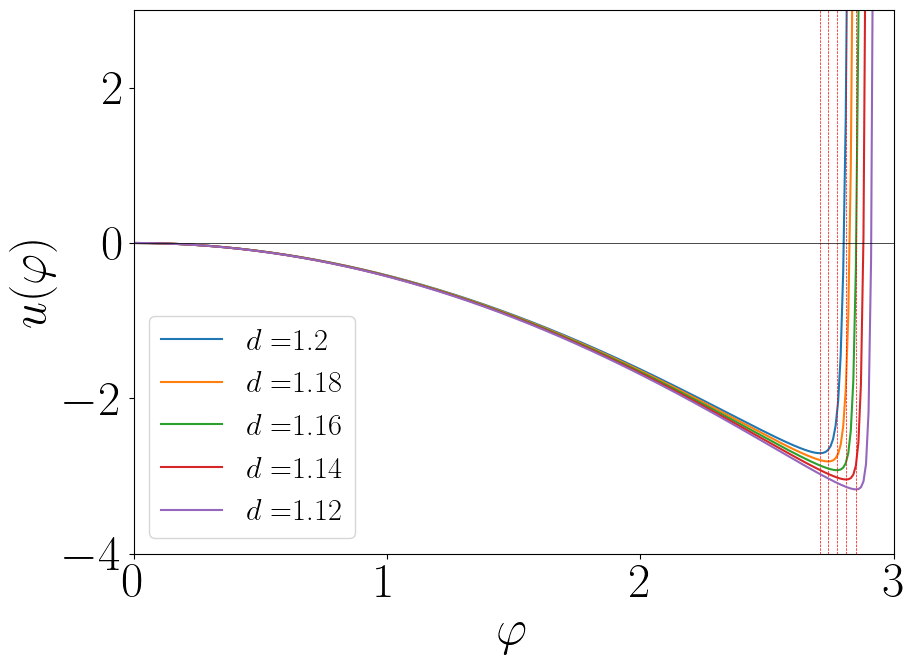}
\caption{Numerical result for the fixed-point effective potential $u(\varphi)$ at the $\partial^2$ order of the derivative expansion when 
approaching the lower critical dimension. This is illustrated for the Exponential regulator with the prefactor $\alpha=1$.The vertical dashed 
lines denote the loci of the minimum of the effective potential for the dimensions shown. The function being symmetric under inversion, we 
only display positive values of the field.
}
    \label{fig:u_typical}
\end{figure}

\begin{figure}
    \centering
    \includegraphics[width=1\linewidth]{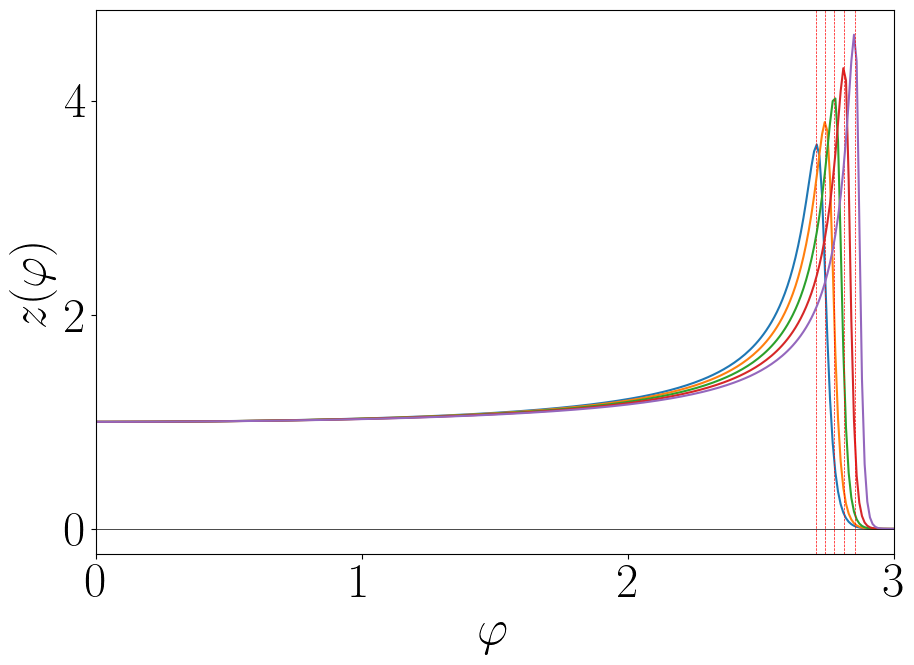}
\caption{Numerical result for the fixed-point for the field renormalization function $z(\varphi)$ at the $\partial^2$ order of the derivative expansion 
when approaching the lower critical dimension. This is illustrated for the Exponential regulator with the prefactor $\alpha=1$. The vertical dashed 
lines denote the loci of the minimum of the effective potential for the dimensions shown. The function being symmetric under inversion, we 
only display positive values of the field. The color code is the same as in Fig.~\ref{fig:u_typical}.
}
    \label{fig:z_typical}
\end{figure}

\subsection{Singular perturbation theory and matching procedure}

In the singular perturbation theory of ordinary differential equations, when a control parameter goes to zero,\cite{KevorkianCole1981, Hunter2004} 
one splits the range of the variable, here the field $\varphi$ from $0$ to $\infty$ (from now on we restrict ourselves to the positive fields, the 
negative range being obtained through the $Z_2$ symmetry), into several regions. In each one of these, a partial solution to the appropriate 
simplified differential equation 
when $\tilde\epsilon\to 0$ is looked for. In the present case we expect three distinct regions, large fields $\varphi \gg \varphi_{\min} \to +\infty$, 
fields of O(1) with $\varphi<\varphi_{\min}$, and a vanishingly narrow boundary (interior) layer near the minimum $\varphi_{\min}$. In the 
former region, the solution of Eqs.~(\ref{eq_ERGEdimensionless_FP}) is exactly known at leading order because one simply neglects the beta 
functions so that one obtains power-law behavior for both  $w$ and $z$, $w(\varphi)\sim \varphi^{1/\tilde\epsilon}$ and 
$z(\varphi)\sim \varphi^{-\eta/\tilde\epsilon}$. 

In the region including $\varphi={\rm O}(1)$ we can neglect the second term in the fixed-point equations and look for a solution of the set
\begin{equation}
\begin{aligned}
\label{eq_ERGEdimensionless_FP_inside}
&0=- d w(\varphi) + \beta^{(d)}_{w}(\varphi;\eta)\\&
0 = \eta  z(\varphi) + \beta^{(d)}_{z}(\varphi;\eta),
\end{aligned}
\end{equation}
where $\eta=2-d=-\beta^{(d)}_{z}(\varphi=0;2-d)$ and $d$ is equal to the ({\it a priori} unknown) lower critical dimension (all of this corresponding 
to $\tilde\epsilon=0$). What makes the present problem more involved than usual singular perturbation theory cases is that not all necessary 
initial conditions of the differential equations in $\varphi=0$ are known. 
The functions being $Z_2$ symmetric, $w'(0)=0$ and $z'(0)=0$, and, as already mentioned, at the $\partial^2$ order of the derivative expansion 
we can choose $z(0)=1$. However, $w(0)=w_0$ is not known; it may still depend on $\tilde\epsilon$, which makes the limit  $\tilde\epsilon\to 0$ 
even more complicated, and it must be determined by some matching procedure with the solution in the other regions. 

The third region is the narrow layer around the minimum, $\vert \varphi-\varphi_{\min}\vert \ll  \varphi_{\min}$. In the layer one defines a rescaled 
variable
\begin{equation}
\label{eq_BLscaling}
x=\frac{\varphi-\varphi_{\min}}{\delta(\tilde{\epsilon})},
\end{equation}
where $\delta(\tilde{\epsilon})\to 0$ is the width of the layer, and this leads to another set of equations at leading order.

All the unknowns should then be fixed by matching the partial solutions in chosen intervals of $\varphi$ where the regions overlap.\cite{KevorkianCole1981, Hunter2004} 
As will be detailed below, this is a subtle issue in the present problem at the $\partial^2$ level.

In the LPA' approximation we could analytically determine the most important features of the fixed-point solution.\cite{lnf23} The solution for the 
effective potential was obtained at leading order in the three regions discussed above. In the boundary layer near the minimum it is described by
\begin{equation}
u(\varphi)= \delta(\tilde{\epsilon}) \upsilon(x)
\end{equation}
with $x$ defined in Eq.~(\ref{eq_BLscaling}) and $\delta(\epsilon)= \tilde{\epsilon}\varphi_{\rm min}$. (Similarly, 
$w(\varphi)=u''(\varphi)=g(x)/\delta(\epsilon)$ with $g(x)=\upsilon''(x)$.) Through a matching procedure between the partial solutions, 
we found that the location of the minimum diverges in a very slow manner as
\begin{equation}
\label{phimin}
\varphi_{\rm min}\sim \sqrt{\ln\left(\frac{1}{\tilde{\epsilon}}\right)}
\end{equation}
while the boundary-layer width shrinks as $\delta(\tilde\epsilon) \sim \tilde{\epsilon} \sqrt{\ln(\frac{1}{\tilde{\epsilon}})}$ when $\tilde\epsilon \to 0$. 
At the same time, the initial value of the square mass function, $w_0=w(\varphi=0)$, approaches the pole of the propagator; see below for more 
detail.

\begin{figure}
    \centering
    \includegraphics[width=1\linewidth]{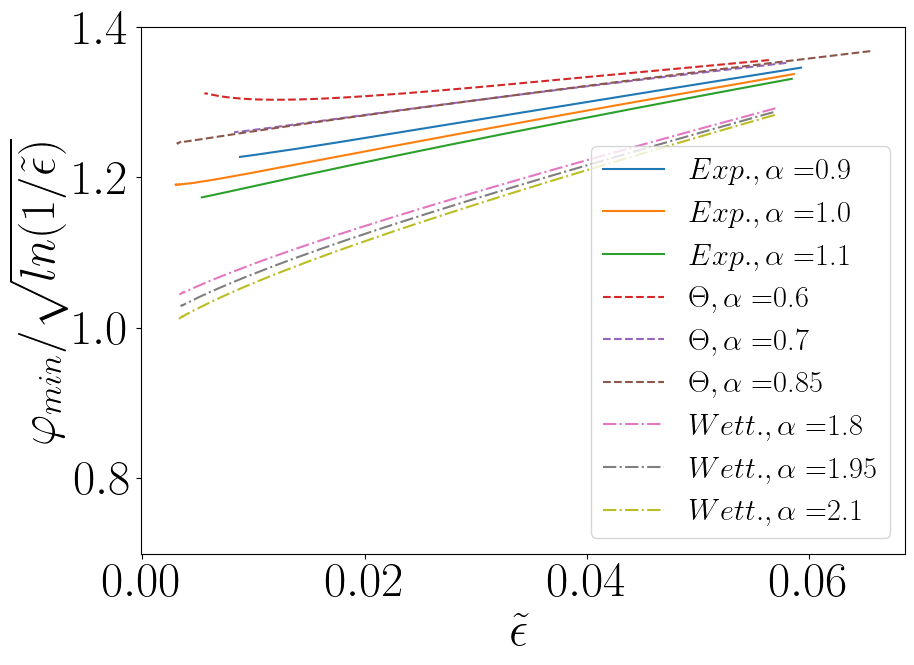}
    \caption{Dependence on $\tilde\epsilon$ of the locus of the minimum of the effective potential, $\varphi_{\rm min}$  rescaled by the expected 
    dependence $\sqrt{\ln(1/\tilde\epsilon)}$ for several choices of regulators at the $\partial^2$ order of the derivative expansion. All the 
    curves seem to go to nonzero values when $\tilde\epsilon\to 0$.
    }
    \label{fig:phim}
\end{figure}

\begin{figure}
    \centering
    \includegraphics[width=0.9\linewidth]{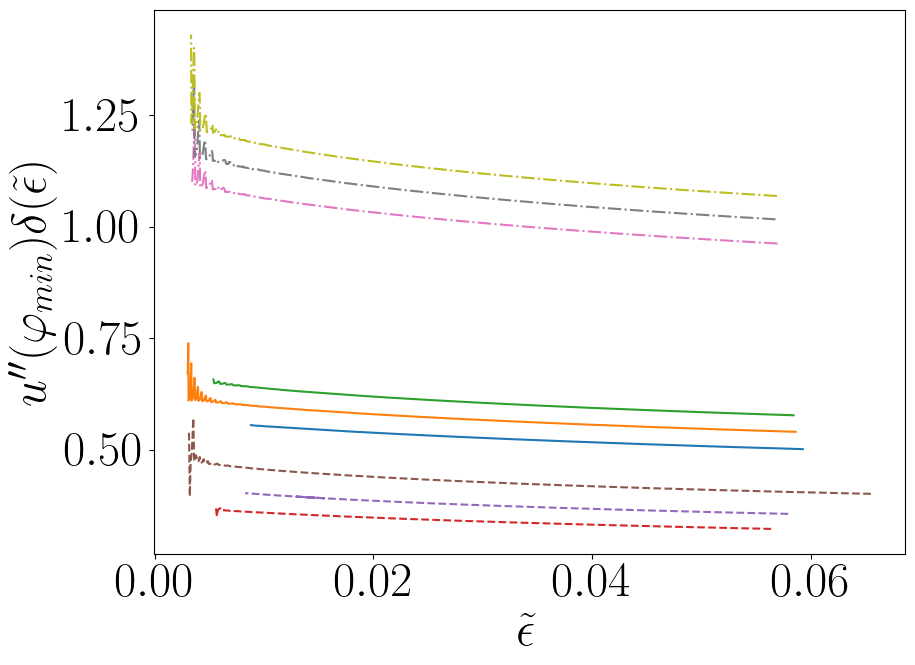}
    \includegraphics[width=0.9\linewidth]{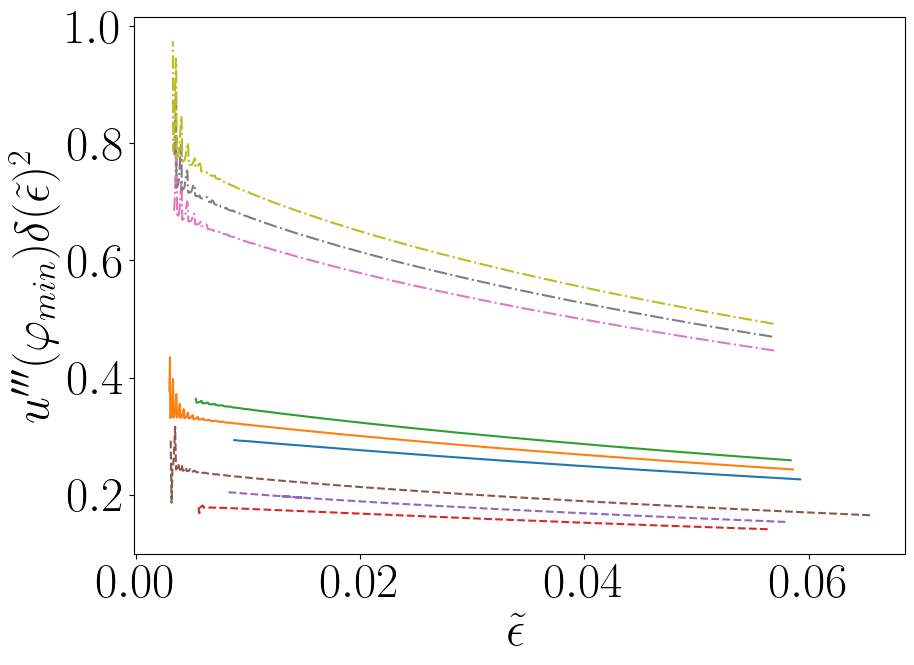}
\caption{Dependence on $\tilde{\epsilon}$ of the second (top) and third (bottom) derivatives of the effective potential at the minimum for 
several choices of regulators (with the same code as in the  legend in Fig.~\ref{fig:phim}) at the $\partial^2$ order of the derivative expansion:  
$u''(\varphi_{\rm min})$ is multiplied by the expected boundary-layer width, $\delta(\tilde{\epsilon})\sim \tilde{\epsilon} \sqrt{\ln(1/\tilde{\epsilon})}$, 
and $u'''(\varphi_{\rm min})$  multiplied by the square of the expected boundary-layer width [compare with Eq.~(\ref{uderBL})]. All the 
curves seem to go to nonzero values when $\tilde\epsilon\to 0$. 
}
    \label{fig:upp+}
\end{figure}

\begin{figure}
    \centering
    \includegraphics[width=0.9\linewidth]{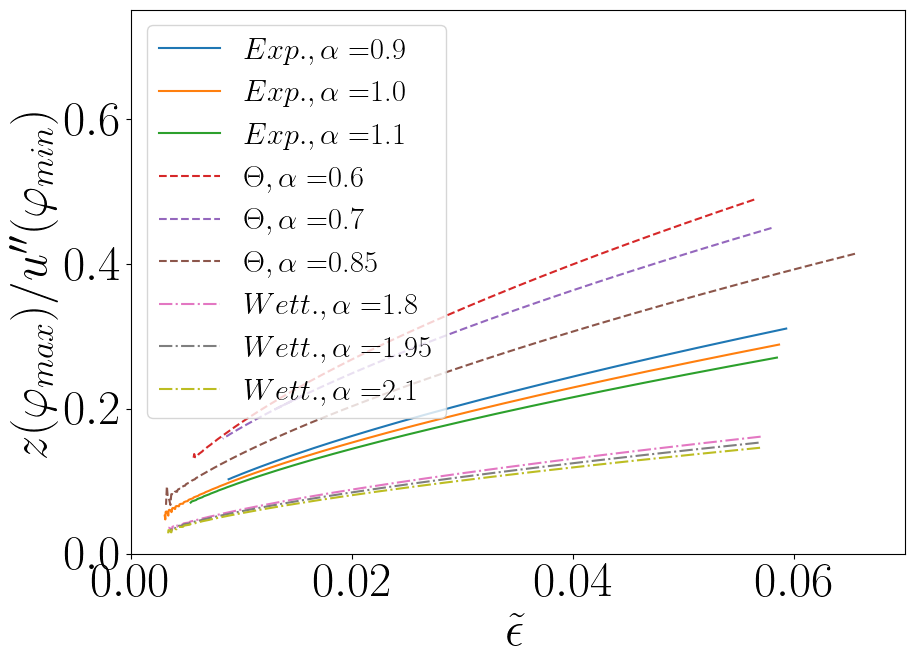}
    \includegraphics[width=0.9\linewidth]{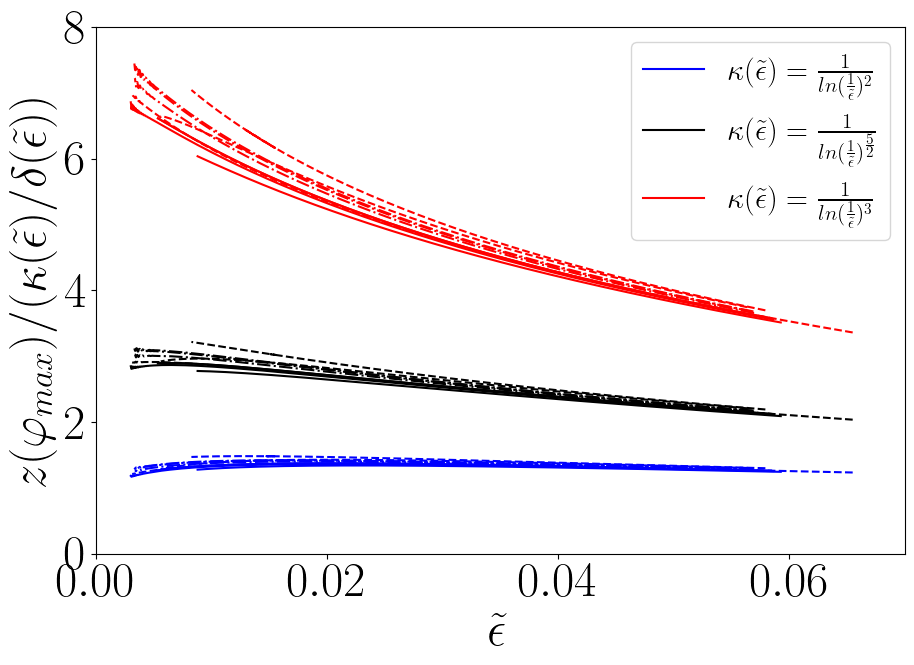}
    \caption{Evidence showing that the maximum of the field renormalization function $z_{\rm max}=z(\varphi_{\rm max})$ grows subdominantly 
    compared to $u''(\varphi_{\rm min})$ when $\tilde\epsilon \to 0$  for several regulator choices (with the same code as in the  legend in 
    Fig.~\ref{fig:phim}). Top: Ratio $z_{\rm max}/u''(\varphi_{\rm min})$ versus $\tilde\epsilon$; a likely extrapolation is that the ratio goes 
    to zero as $\tilde\epsilon\to 0$. Bottom: Same data further divided by a factor $\kappa(\tilde{\epsilon})\to 0$ and illustrated with the functional 
    form $\kappa(\tilde{\epsilon})\propto [\ln(\frac{1}{\tilde{\epsilon}})]^{-\mu}$, with $\mu=2, 2.5, 3$; $\mu\approx 2-2.5$ appears 
    to best collapse the data.
}
    \label{fig:zoupp+}
\end{figure}

\subsection{Numerical evidence for a boundary-layer scenario at $\partial^2$ order}
\label{sub_numerical}

We wish to check whether the same scenario as at the LPA' applies at the $\partial^2$ order. To this end we have numerically determined the 
fixed-point functions for several IR regulators and solved the set of ordinary nonlinear differential equations for $u(\varphi)$ and $z(\varphi)$ 
down to the lowest dimensions approaching the (unknown) lower critical dimension for which numerically reliable solutions can be obtained. 
(This corresponds to small but nonzero values of $\tilde{\epsilon}$.) The results is illustrated in Fig.~\ref{fig:u_typical} for $u(\varphi)$ and in  
Fig.~\ref{fig:z_typical} for $z(\varphi)$. The development of a shrinking layer around the minimum of the potential is clearly visible in both 
functions.

We have next investigated the characteristics of this emerging layer. The results are presented in terms of $\tilde{\epsilon}$, 
which is a direct measure of the relative distance to the lower critical dimension and is the small parameter that naturally arises in the 
fixed-point FRG equations. Contrary to the exact description in which the lower critical dimension is known to be $d_{\rm lc}=1$, and one can 
therefore use $d-1$ as the relative distance to the lower critical dimension as well, the value of $d_{\rm lc}$ is not {\it a priori} known in the 
present approximation scheme and can furthermore vary with the choice of the IR cutoff function (see above). The plots characterizing the 
approach to the lower critical dimension are thus only meaningfully displayed as a function of $\tilde{\epsilon}$ (or, equivalently, of the dimension 
of the field $D_\phi$). The (numerical) correspondence between $D_\phi$ and the dimension $d$ will be shown later on 
in Sec.~\ref{sub_extrapolations}).

We plot in Fig.~\ref{fig:phim} the minimum of the potential according to the 
same scaling with $\tilde\epsilon$ as in the LPA': see Eq.~(\ref{phimin}). The figure convincingly shows that the dependence 
$\varphi_{\rm min}\sim \sqrt{\ln(\frac{1}{\tilde{\epsilon}})}$ is indeed observed. Another property resulting from the boundary-layer form is that the 
second and higher-order derivatives of the effective potential evaluated at the minimum diverge with increasing powers of $\delta(\tilde{\epsilon})$ 
as
\begin{equation}
\label{uderBL}
u^{(p)}(\varphi_{\rm min})\propto\frac{1}{\delta(\tilde{\epsilon})^{p-1}}.
\end{equation}
The outcome is displayed in Fig.~\ref{fig:upp+} for the second and third derivatives of the potential and it indeed supports 
the boundary-layer form with the same dependence on $\tilde{\epsilon}$ as at the LPA' approximation.

We now turn to the boundary layer in the field renormalization function. The emergence of a boundary layer is clearly seen from 
Fig.~\ref{fig:z_typical}, with the location of maximum in $z(\varphi)$ being very close to that of the minimum in $u(\varphi)$. The maximum in 
$z(\varphi)$ grows when $\tilde{\epsilon}$ decreases, but it appears to do so more slowly than $w(\varphi_{\rm min})=u''(\varphi_{\rm min})$. 
Indeed, we display the ratio $z(\varphi_{\rm max})/u''(\varphi_{\rm min})$ in Fig.~\ref{fig:zoupp+}, and the data points to a ratio going to zero as 
$\tilde{\epsilon}\to 0$. We characterize this subdominance of $z(\varphi)$ compared to $u''(\varphi)\equiv w(\varphi)$ near the minimum by the 
factor $\kappa(\tilde{\epsilon})$, i.e., $z(\varphi\approx  \varphi_{\rm min})\sim \kappa(\tilde{\epsilon}) w(\varphi_{\rm min})$, 
 and we empirically find that this factor goes to zero as $\tilde{\epsilon}\to 0$ rather slowly, an appropriate functional form being
\begin{equation}
\kappa(\tilde{\epsilon})\propto \frac 1{[\ln(\frac{1}{\tilde{\epsilon}})]^{\mu}},
\end{equation} 
with $\mu\approx 2.5$ giving the best data collapse; see Fig.~\ref{fig:zoupp+}.

\section{Boundary layer solution at $\partial^2$ order}
\label{sec:BL}

Based on the above numerical results and on our previous work,\cite{lnf23} we look for a boundary-layer solution near the minimum of the 
potential in the form
\begin{equation}
    \label{bl_ansatz_w}
    w(\varphi)\equiv u''(\varphi)= \frac{g(x)}{\delta(\tilde{\epsilon})}
\end{equation}
with $x$ given in Eq.~(\ref{eq_BLscaling}) and 
\begin{equation}
    \label{bl_ansatz_z}
    z(\varphi)= \frac{\kappa(\tilde{\epsilon})}{\delta(\tilde{\epsilon})}\zeta(x),
\end{equation}
where, by balancing the first two terms of the fixed-point equations for $w$ and $z$, we can choose  
$\delta(\tilde{\epsilon})=\tilde{\epsilon}\varphi_{\rm min}(\tilde{\epsilon})$ and where $\varphi_{\rm min}(\tilde{\epsilon})\to \infty $ and 
$\kappa(\tilde{\epsilon})\to 0$ in some yet to be determined way as $\tilde{\epsilon}\to 0$. Note that $\kappa(\tilde{\epsilon})$ is defined up to a 
multiplicative constant and we can then choose $\zeta(0)=1$.

The ansatz in Eq.~(\ref{bl_ansatz_z}) is compatible with the numerical findings displayed in Fig.~\ref{fig:zoupp+}, and we give additional arguments 
in Appendix~\ref{app:marginal_cases}. There, we study two alternative possibilities: 1) $\kappa(\tilde{\epsilon})/\delta(\tilde{\epsilon})={\rm O}(1)$, 
so that $z={\rm O}(1)$ in the boundary layer; 2) $\kappa(\tilde{\epsilon})={\rm O}(1)$, so that $z$ and $w$ scale in the same way. For these 
two cases we find that there is no possible matching with a physically reasonable solution in the interior interval where $\varphi={\rm O}(1)$.

With the above expressions for $w(\varphi)$ and $z(\varphi)$ we can derive the fixed-point equations in the boundary layer in terms of 
the variable $x$. Indeed, the dimensionless propagator that appears in the beta functions (see Appendix~\ref{appx_flows}) reads
\begin{equation}
\begin{aligned}
    p(y;\varphi)&=\frac{1}{y [r(y) +z(\varphi)] + w(\varphi)} \\&
    \sim \frac{\delta(\tilde{\epsilon})}{g(x)+ \kappa(\tilde{\epsilon})\zeta(x)y+\delta(\tilde{\epsilon})y r(y)}
\end{aligned}
\end{equation}
where as before $y$ is the dimensionless square momentum. As $\delta(\tilde{\epsilon})\to 0$ and $\kappa(\tilde{\epsilon})\to 0$ when 
$\tilde{\epsilon}\to 0$, we can then simplify the beta functions in the boundary layer. After first rescaling the field $\varphi$ by a factor 
$\sqrt{d/(2 v_d \alpha A_d)}$, where $A_d$ is a regulator-dependent constant, $A_d=d \int_0^{\infty}dy y^{d/2} [r(y)/\alpha]$, we obtain the 
following fixed-point equations at leading order in $\kappa(\tilde{\epsilon})$, $\delta(\tilde{\epsilon})$, and $\tilde{\epsilon}$ (note that we anticipate 
that $\kappa(\tilde{\epsilon})\gg\delta(\tilde{\epsilon})\gg\tilde{\epsilon}$),
\begin{equation}
\begin{aligned}
\label{eq_g-zeta_BL}
&0=-g(x)+g'(x)+ \partial^2_x\left(\frac{1}{g(x)}\right)\\&
0=\frac{(2-d)}d \zeta(x) + \zeta'(x) - \Bigg [ \frac{\zeta''(x)}{g(x)^2}+ 4\frac{\zeta'(x)}{g(x)}\partial_x\left(\frac{1}{g(x)}\right) \\&
+ 2\zeta(x)\left(\partial_x\left(\frac{1}{g(x)}\right)\right)^2 \Bigg],
\end{aligned}
\end{equation}
where we have used that $\eta=2-d+O(\tilde{\epsilon})$. At the leading order considered here, $d$ should be taken equal to $d_{\rm lc}$. 
(Note that the equation for $\zeta(x)$ is linear in $\zeta(x)$ so that, as already stated, one can choose $\zeta(0)=1$, which simply 
amounts to a redefinition of $\kappa(\tilde\epsilon)$ by a constant.) One notices that any explicit reference to the IR cutoff 
function has now disappeared from the boundary-layer equations. However, one should keep in mind that some of the constants, e.g., 
the functions and their derivative in $x=0$, are unknown and must be determined by a matching procedure. As described below, when 
matching takes place with a solution in another region which itself depends on the IR regulator, one finds that the global solution including 
the boundary-layer piece also depends on the regulator choice.

Similarly, the equation for the location of the minimum [see Eq.~(\ref{eq_ERGEdimensionless_min_FP})] can be written as
\begin{equation}
\label{eq_minimum_BL}
0=g(0)-\frac{g'(0)}{g(0)^2}.
\end{equation}
Note that the equation for $g(x)$ is independent from $\zeta(x)$ 
and is the same as that obtained in the LPA' approximation.\cite{lnf23} 
\\

To solve the boundary-layer equations we introduce an auxiliary function 
\begin{equation}
\label{eq:aux_Y}
    X(x)=\partial_x\left(\frac{1}{g(x)}\right)+ g(x),
\end{equation}
which satisfies $X'(x)=g(x)$ [because of the first equation in  (\ref{eq_g-zeta_BL})] and $X(0)=0$ [because of Eq.~(\ref{eq_minimum_BL})]. 
One can then express $x$ as a function of $X$ and define
\begin{equation}
\begin{aligned}
\label{eq:defa-b}
&a(X)\equiv\frac{1}{g(x)} \\&
b(X)\equiv\frac{\zeta(x)}{g(x)^2}.
\end{aligned}
\end{equation}

In terms of these new functions $a(X)$ and $b(X)$, the equations in (\ref{eq_g-zeta_BL}) can be rewritten as
\begin{equation}
\begin{aligned}
\label{eq:bl_fpe_a-b}
&a'(X)-X a(X) =-1,\\&
b''(X)-X b'(X) - \frac{2+d}{d} b(X) =0.
\end{aligned}
\end{equation}

The solution of the first equation is easily obtained as
\begin{equation}
 a(X)= e^{\frac{X^2}{2}}\left(a_0-\int^{X}_{0}dX'\,e^{-\frac{X'^2}{2}}\right). 
\end{equation}
Interestingly, this form allows us to proceed to the matching with the large-field solution when $\varphi\to \infty$. The latter is simply 
obtained from the dimensional part of the fixed-point equations, Eq.~(\ref{eq_ERGEdimensionless_FP}), which leads to 
$w(\varphi)\sim \varphi^{1/\tilde\epsilon}\to +\infty$ (see above). Choosing a matching region such that $1/\tilde\epsilon\gg x\gg 1$, 
this imposes that $g(x)\sim e^x$ at large $x$. As a result $X\to +\infty$ when $x\to +\infty$ and $a(X\to+\infty)\to 0$. This can only be 
satisfied if the constant $a_0$ appearing in the above solution is equal to $\int^{\infty}_{0}dX'\,\exp(-X'^2/2)=\sqrt{\pi/2}$, so that 
\begin{equation}
\label{eq_a(X)} 
a(X)= e^{\frac{X^2}{2}}\int_{X}^{\infty}dX'\,e^{-\frac{X'^2}{2}}. 
\end{equation}
This is the same solution as found in the LPA' approximation.\cite{lnf23} The solution for $g(x)$ is then obtained by combining the above 
with Eqs.~(\ref{eq:aux_Y}) and (\ref{eq:defa-b}). From the value in $X=x=0$, one immediately derives that $g(0)=\sqrt{2/\pi}$ and, from 
Eq.~(\ref{eq_minimum_BL}), that $g'(0)=(2/\pi)^{3/2}$. The function $g(x)$ is therefore fully determined but to obtain the corresponding 
$w(\varphi)$ one further needs $\delta(\tilde\epsilon)$ and therefore $\varphi_{\rm min}(\tilde\epsilon)$ .

We next investigate the solution for the function $b(X)$. After introducing the parameter
\begin{equation}
q=\frac{d+2}{d},
\end{equation} 
we can obtain the solution in the form 
\begin{equation}
\label{eq:bX}
b(X)=H_{-q}\left(\frac{X}{\sqrt{2}}\right)\, \hat b + {}_1F_1\left(\frac{q}{2},\frac{1}{2},\frac{X^2}{2}\right)\,\tilde b,
\end{equation}
where $\hat b$ and $\tilde b$ are two constants to be determined, $H_{-q}$ is the Hermite function generalized to negative real degrees 
[note that $a(X)$ in Eq.~(\ref{eq_a(X)}) can also be expressed as $H_{-1}(X/\sqrt 2)$], and ${}_1F_1$ is Kummer's confluent 
hypergeometric function. We consider the matching with the large field region where $z(\varphi)\sim \varphi^{-(2-d)/(d\tilde\epsilon)}$ 
(see above); when $1/\tilde\epsilon\gg x\gg 1$, this implies that $\zeta(x)$ goes to zero as $\exp[-(2/d-1)x]$. 
From the known asymptotic behavior of the Hermite and hypergeometric functions when $X\gg 1$, we then conclude that $\tilde b=0$ while  
$\hat b> 0$ is constrained by the condition $\zeta(0)=1$. By using $g(0)=\sqrt{2/\pi}$, $H_{-q}(0)=2^{-q}\sqrt \pi/\Gamma((1+q)/2)$, and 
Eq.~(\ref{eq:defa-b}) in $X=0$, we obtain that $\hat b=2^{q-1}\sqrt \pi \Gamma((1+q)/2)$. 

By expressing the derivative with respect to $x$ in two different ways, i.e., by deriving the right-hand side of the second equation in 
Eq.~(\ref{eq:defa-b}) or by deriving the solution for $b(X)$ and using that $X'(x)=g(x)$, we arrive at 
\begin{equation}
\begin{aligned}
\label{match_de2}
\frac{\zeta^{\prime}(0)}{g(0)\zeta(0)}=2\sqrt{\frac{2}{\pi}}-\sqrt{2}\frac{\Gamma\left(\frac{(1+q)}2\right)}{\Gamma(\frac{q}2)},
\end{aligned}
\end{equation}
so that, since $g(0)=\sqrt{\pi/2}$ and $q=(d+2)/d$,
\begin{equation}
\begin{aligned}
\label{match_de2bis}
\frac{\zeta^{\prime}(0)}{\zeta(0)}=2\left [1-\frac{\Gamma(\frac 32)\Gamma(1+\frac{1}{d})}{\Gamma(\frac{1}{2}+\frac{1}{d})}\right ],
\end{aligned}
\end{equation}
where $d\equiv d_{\rm lc}$.

While the solution for $g(x)$ is fully determined by the matching procedure with the large-field region, this is not the case for the function 
$\zeta(x)$ which is still parametrized by $d_{\rm lc}$. Equivalently, this means that the value of the lower critical dimension is not simply 
determined by considering the boundary-layer solution and the matching with the large-field region, as it was in the LPA' approximation.\cite{lnf23}
(In the LPA' the renormalization prescription for the field renormalization function, which is otherwise taken as independent of the field,  
is chosen at the minimum of the potential, $z(\varphi_{\rm min})=1$, and therefore within the boundary layer.) 

\begin{figure}
    \centering
    \includegraphics[width=1\linewidth]{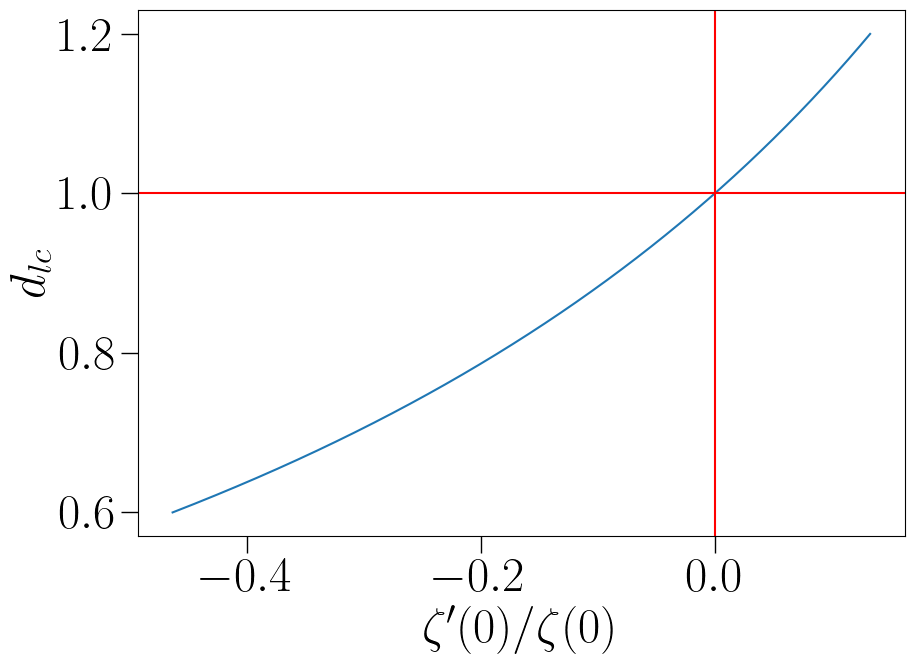}
    \caption{Relation between the lower critical dimension $d_{\rm lc}$ and the ratio $\zeta^{\prime}(0)/\zeta(0)$ parametrizing the fixed-point 
    solution for the field renormalization function in the boundary layer, as obtained from Eq.~(\ref{match_de2bis}). 
    }
    \label{fig:zp_vs_d}
\end{figure}

We plot in Fig.~\ref{fig:zp_vs_d} the monotonic relation between $d_{\rm lc}$ and the parameters of the fixed-point solution for the field 
renormalization function in the boundary layer $\zeta'(0)/\zeta(0)$ given in Eq.~(\ref{match_de2bis}). 
Interestingly, the exact value $d_{\rm lc}=1$ is recovered if the minimum of the potential corresponds to an extremum of $z$, i.e.,
$\zeta'(0)=0$. Note, however, that for the full solution over the whole field range, it is the matching to the inner solution that should 
determine the actual value of $\zeta'(0)$, as we proceed to explain. 

To obtain a full solution over the whole field interval $[0,\infty)$ one needs to solve the equation outside the boundary layer in the region including 
fields of O(1) and to match this partial solution with the boundary-layer solution in a region where $1\ll \varphi \ll \varphi_{\rm min}$ (as we 
anticipate that $\varphi_{\rm min}$ diverges when $\tilde\epsilon\to 0$, see Fig.~\ref{fig:phim}) and $x \to -\infty$, e.g., 
$\varphi_{\rm min}-\varphi={\rm O}((\tilde\epsilon \varphi_{\rm min})^a)$ with $0<a<1$.
\\

When $x \to -\infty$, one easily finds that 
\begin{equation}
\label{eq_g(x)_matching}
g(x) \sim \frac 1{\sqrt 2 \vert x\vert \sqrt{\ln( \vert x\vert)}} \to 0
\end{equation}
and 
\begin{equation}
\label{eq_zeta(x)_matching}
\zeta(x) \sim \frac {[\ln( \vert x\vert)]^{\frac{2-d}{2d}}}{\vert x\vert} \to 0,
\end{equation}
so that in the matching region $w(\varphi)\sim {\rm O}((\tilde\epsilon \varphi_{\rm min})^{-a})$ is much larger than $1$ and 
$z(\varphi) \sim \kappa(\tilde\epsilon){\rm O}((\tilde\epsilon \varphi_{\rm min})^{1-2a})$ is also very large if $a>1/2$ 
and $\kappa$ is only slowly going to zero.

Matching with the solution of the equation derived for $\varphi$=O(1) is thus possible only if the latter can reach 
field values $\varphi\gtrsim \varphi_{\rm min}$ and if in the matching region both the square mass $w$ and the field renormalization function $z$ 
are very large when $\tilde\epsilon \to 0$. As we will see, these turn out to be very stringent constraints at the $\partial^2$ order of the derivative 
expansion. This is expected because one still needs to determine several quantities: $d_{\rm lc}$ or $\zeta'(0)$, $\delta(\tilde\epsilon)$ 
or equivalently $\varphi_{\rm min}(\tilde\epsilon)$, $\kappa(\tilde\epsilon)$, and the initial condition $w_0$ in $\varphi=0$.
\\

\section{Matching procedure at $\partial^2$ order}
\label{sec:matching_de2}

\subsection{Summary of the procedure in the LPA' approximation}

In the LPA' approximation where the field renormalization is a scale-dependent but field-independent parameter fixed by $z(\varphi_{\rm min})=1$, 
the only two remaining unknowns are $\varphi_{\rm min}(\tilde\epsilon)$ and $w_0$ and they should be fixed (at leading order) by finding a 
solution of the $\varphi$=O(1) region that matches the boundary-layer solution when $\varphi$ approaches $\varphi_{\rm min}$.\cite{lnf23} 
The equation to be solved is the first one of Eq.~(\ref{eq_ERGEdimensionless_FP_inside}), which explicitly reads
\begin{equation}
\label{eq:fp_w}
w(\varphi)= \frac{2v_d}d \partial^2_\varphi \ell^{(d)}_0(w(\varphi);\eta=2-d),
\end{equation}
where $\ell^{(d)}_0$ is a threshold function involving the IR regulator and is defined as (see Appendix~\ref{appx_flows})
\begin{equation}
\ell^{(d)}_0(w(\varphi);\eta)=-\frac 12\int_0^\infty dy y^{\frac d2}\, \frac{\eta r(y)+2y r'(y)}{y[z+r(y)]+w}
\end{equation}
with $z\equiv 1$; $d=d_{\rm lc}$ is known from the boundary-layer solution which is itself fully determined by the matching with 
 the large-field region where the beta functions vanish; see above and [\onlinecite{lnf23}].

The function $\ell^{(d)}_0(w;\eta=2-d)$ being monotonically decreasing with $w$, Eq.~(\ref{eq:fp_w}) can be turned into the equation of motion 
for an anharmonic oscillator, $\varphi$ playing the role of time, from which one derives the key properties:\cite{lnf23}
\begin{itemize}
\item All solutions $w(\varphi)$ are periodic with a half period $\varphi_*$ that depends on the initial condition $w_0$.
\item The half-period $\varphi_*$ can become very large and $w_*=w(\varphi_*)$ very large as well, both conditions being required for 
matching the boundary-layer solution with $\varphi_{\rm min}\gg 1$ and $w(\varphi_{\rm min})\sim 1/\delta(\tilde\epsilon)\gg1$, provided 
$w_0$ approaches closely enough the pole $w_P$ of the propagator $p(y;\varphi)=[y(1+r(y))+w]^{-1}$: $w_0=w_P +\rho(\tilde\epsilon)$, 
where $\rho(\tilde\epsilon\to 0)\to 0^+$ in a manner that depends on the detailed form of the IR cutoff function $r(y)$.
\item When $\varphi_*$ and $w_*$ are both very large, one can show that $\varphi_*\sim \sqrt{\ln w_*}$.
\end{itemize}

A global solution of the fixed-point equation requires one to match the solution just obtained with that in the boundary layer. As already 
stressed it however represents an unusual form of singular perturbation theory because the initial value $w_0$ is not known and moreover 
still depends on the small parameter $\tilde\epsilon$ (in an equation which is otherwise considered at $\tilde\epsilon=0$). The matching 
needs $\varphi_*\gtrsim \varphi_{\rm min}$ and $w_*\gtrsim w(\varphi_{\rm min})$ and the overlap region where the two solutions should  
coincide at leading order is such that all these quantities are very large (see also above). In this limit it is easily seen that the equations 
derived with $\varphi={\rm O}(1)$ and that in the boundary layer coincide. The equations are the same but the boundary 
conditions are $w_*$ in $\varphi_*$ for the former and $w(\varphi_{\rm min})=1/(\tilde\epsilon\varphi_{\rm min})$ in $\varphi_{\rm min}$ for 
the latter. Matching is then automatically guaranteed by taking
\begin{equation}
w_*\sim w(\varphi_{\rm min})=1/(\tilde\epsilon\varphi_{\rm min})\;\; {\rm and}\;\; \varphi_*\sim \varphi_{\rm min}.
\end{equation}
With the property that $\varphi_*\sim \sqrt{\ln w_*}$ this immediately leads to $\varphi_{\rm min}\sim \sqrt{\ln[1/(\tilde\epsilon\varphi_{\rm min})]}$, 
hence $\varphi_{\rm min}\sim \sqrt{\ln(1/\tilde\epsilon)}$ at dominant order.

We conclude this recap with three remarks. First, this represents a weak form of matching because only orders of magnitude are matched. 
Going beyond this would be very involved. Second, the periodic solution of the $\tilde\epsilon=0$ equation has itself a boundary layer around 
the half period $\varphi_*$ with a width $1/w_*$ that goes to zero as $w_0$ approaches the pole of the propagator 
and $w_*$ diverges. In the matching region, the solution of the (physical) boundary layer and that of the boundary layer of the periodic 
solution coincide, being both given by Eq.~(\ref{eq_g(x)_matching}). Finally, the fact that $w_0$ approaches the pole of the propagator is 
expected on physical ground for the lower critical dimension where the critical and the zero-temperature ordered fixed points merge.\cite{lnf23}

\subsection{Scenario for matching at $\partial^2$ order}
\label{sub_scenarioDE2}

At the $\partial^2$ order singular perturbation theory requires us to consider the equations in Eq. (\ref{eq_ERGEdimensionless_min_FP}) when 
dropping the second terms in $\tilde\epsilon \varphi\partial_\varphi$ and neglecting terms of order $\tilde\epsilon$. As already stressed,  
$w_0$ is unknown, but the matching with the boundary layer solution should fix $\delta(\tilde\epsilon)$ [or equivalently 
$\varphi_{\rm min}(\tilde\epsilon)$], $\kappa(\tilde\epsilon)$, and  $d_{\rm lc}$ [or $\zeta'(0)$]; see above. 

We again anticipate that a requirement for a possible matching is that $w_0$ approaches the pole $w_P$ of the propagator. 
We find, however, that this is very hard to achieve while (numerically) getting solutions that are physically acceptable. We have been unable to 
obtain a rigorous resolution of the problem. Actually, the fine-tuning of the value of $d_{\rm lc}$ that provides a proper matching appears so 
demanding (with severe  numerical instabilities developing when $w_0\to w_P$) that our procedures are unable to find a definite answer. 
(Note that, as a result, we cannot exclude the possibility that the fixed-point equations stop having a solution when $\tilde\epsilon$ is at and 
below a very small but nonzero threshold, i.e., slightly before reaching $d_{\rm lc}$.)

Lacking strong enough analytical and numerical control of the solutions, we thus make a tentative set of hypotheses based on the numerical 
findings shown in Sec.~\ref{sec_nonuniform} and the scenario found at the LPA' order:
\begin{itemize}
\item The relevant solution of the set of equations obtained by dropping the $\tilde\epsilon \varphi\partial_\varphi$ terms is periodic with a 
half period $\varphi_*$ that depends on the initial condition $w_0$. It is denoted by ($\overline w(\varphi)$, $\overline z(\varphi)$).
\item When $w_0$ approaches the pole $w_P$ of the propagator, the half-period $\varphi_*$, as well as $w_*=\overline w(\varphi_*)$ and 
$z_*=\overline z(\varphi_*)$ can become very large, with nonetheless $z_* \ll w_*$. 
\item The existence of a periodic solution when $w_0\to w_P$ (and $\tilde\epsilon\to 0$) requires the selection of a unique, and of course 
regulator-dependent, value $d_{\rm lc}$.
\item Furthermore, when $\varphi_*$ and $w_*$ are both very large, they are related by $\varphi_*\sim \sqrt{\ln w_*}$.
\end{itemize}

Before giving evidence supporting the above hypotheses, we explore their consequences. First, one easily realizes that in the matching 
region between the periodic and the boundary-layer solutions, where $1\ll \varphi\ll \varphi_*, \varphi_{\rm min}$, $1\ll w\ll w_*, w(\varphi_{\rm min})$, 
$1\ll z\ll z_*, z(\varphi_{\rm min})$ with $z\ll w$, the equations are identical because the terms in $\tilde\epsilon \varphi\partial_\varphi$ are 
always negligible. The difference is of course in the boundary values that are taken in $\varphi_*$ for the periodic solution 
$(\overline w(\varphi), \overline z(\varphi))$  and in $\varphi_{\rm min}$ for the boundary-layer solution. Matching between the two types of solutions 
is then automatically achieved if $\varphi_* \sim \varphi_{\rm min}$, $w_* \sim w(\varphi_{\rm min})$, and $z_* \sim z(\varphi_{\rm min})$. With 
the last assumption that $\varphi_*\sim \sqrt{\ln w_*}$, this immediately yields
\begin{equation}
\varphi_{\rm min} \sim \sqrt{\ln\left(\frac 1{\tilde\epsilon}\right)}\;\; {\rm and}\;\; \delta(\tilde\epsilon)\sim \tilde\epsilon\sqrt{\ln\left(\frac 1{\tilde\epsilon}\right)},
\end{equation}
as at the LPA' level of approximation and as well confirmed numerically (see Sec.~\ref{sub_numerical}). However, in the absence of 
more detailed information concerning the actual analytical form of the periodic solution $\overline z(\varphi)$, we are unable to determine the 
scaling of $\kappa(\tilde\epsilon)$.

\begin{figure}[h!]
  \begin{center}
 \includegraphics[width=1\linewidth]{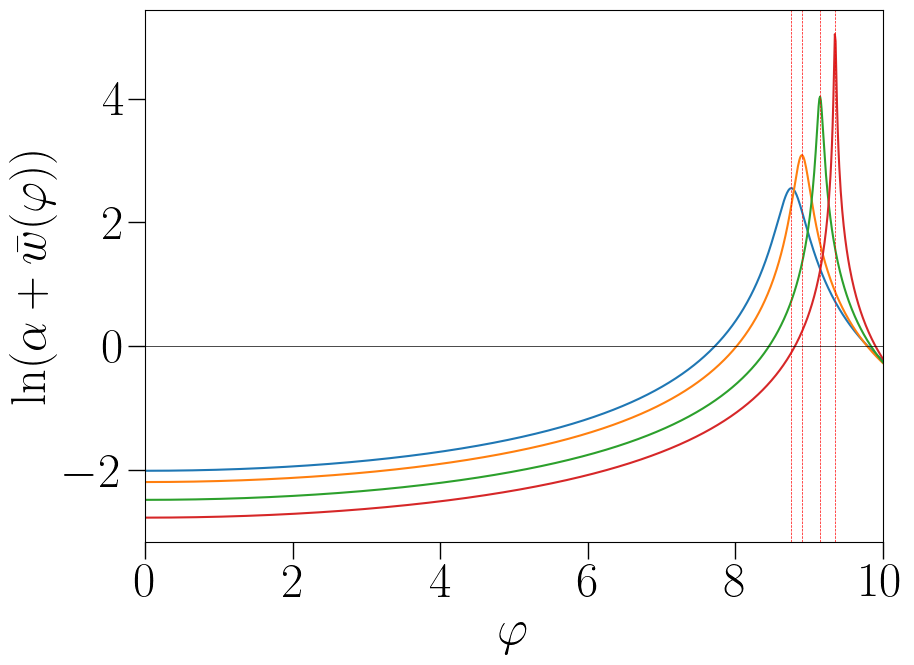}
    \caption{Periodic solution of Eq.~(\ref{eq_ERGEdimensionless_FP_inside_periodic}): $\ln(\alpha +\overline w(\varphi))$ versus $\varphi$ 
    for $d=0.8$  and the Litim/Theta regulator with $\alpha=0.7$. The different curves represent different initial conditions of $w_0>w_P=-\alpha$. 
    As $w_0$ approaches closer to the pole of the propagator $w_P$, the half-period $\varphi_*$ at which $\overline w$ is maximum shifts to higher 
    values while $w_*=\overline w(\varphi_*)$ rapidly increases. We only show the periodic function in a range of field between $0$ and 
    somewhat above the half-period which is marked by a vertical dashed line.
    }
    \label{fig:chi_per}
  \end{center}
\end{figure}

\begin{figure}[h!]
  \begin{center}
 \includegraphics[width=1\linewidth]{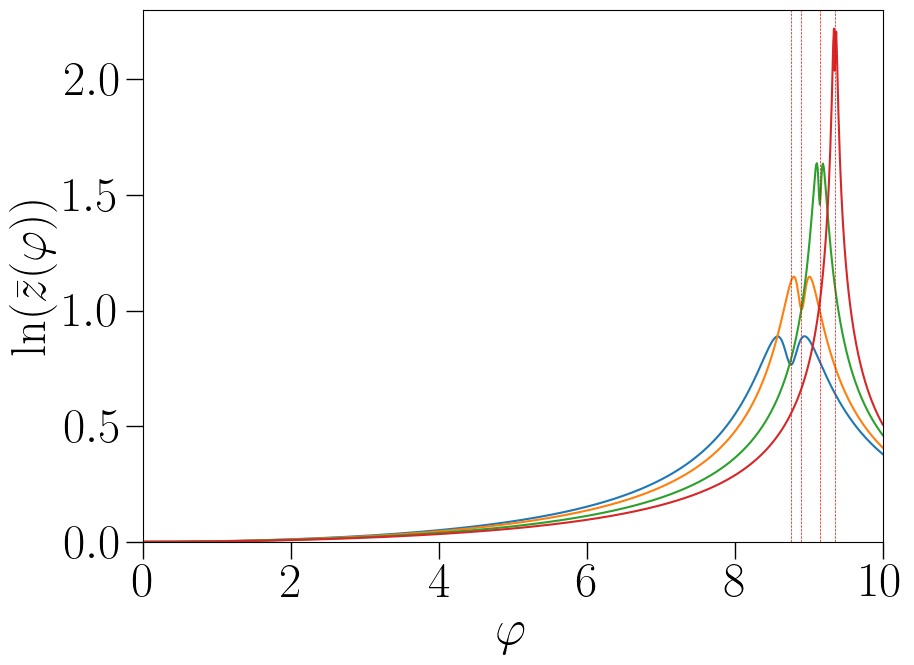}
    \caption{Periodic solution of Eq.~(\ref{eq_ERGEdimensionless_FP_inside_periodic}): $\ln(\overline z(\varphi))$ versus $\varphi$ 
    for $d=0.8$  and the Litim/Theta regulator with $\alpha=0.7$. The curves correspond to the same initial conditions $w_0>w_P=-\alpha$ as in 
    Fig.~\ref{fig:chi_per} and the half-period $\varphi_*$ is in each case the same as for $\overline w$. Note that $\overline z$ has a 
    minimum in $\varphi_*$ and first passes through a maximum for a value which is very close to $\varphi_*$. The value $z_*=\overline z(\varphi_*)$ 
    increases as $w_0$ approaches closer to the pole of the propagator $w_P=-\alpha$.  We only show the periodic function in a range of field 
    between $0$ and somewhat above the half-period which is marked by a vertical dashed line.
    }
    \label{fig:z_per}
  \end{center}
\end{figure}

\subsection{Searching for periodic solutions of the equations derived for $\varphi={\rm O}(1)$}
\label{sub_periodic}

\begin{figure}[h!]
  \begin{center}
 \includegraphics[width=1.0\linewidth]{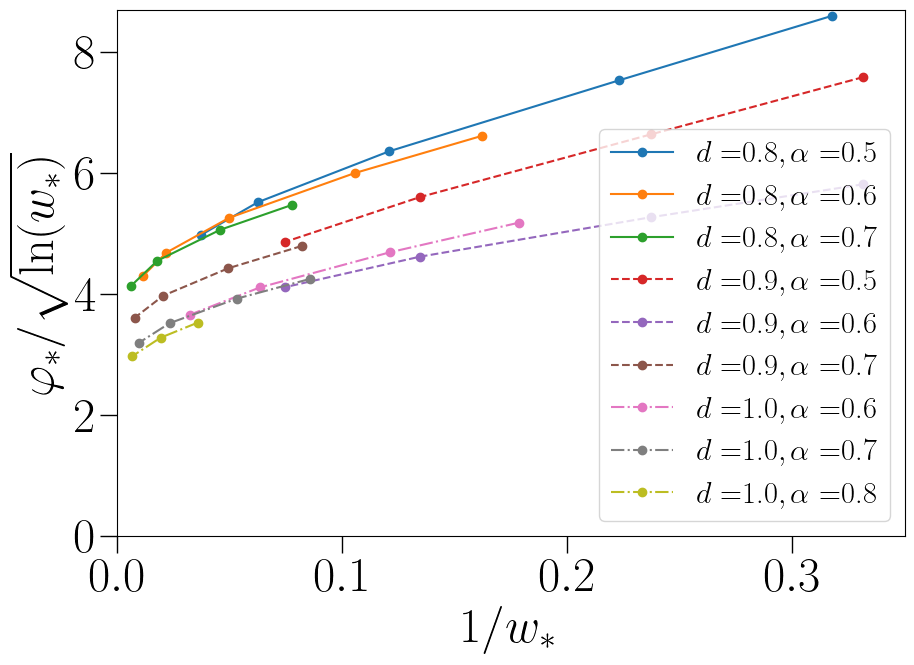}
  \includegraphics[width=1.0\linewidth]{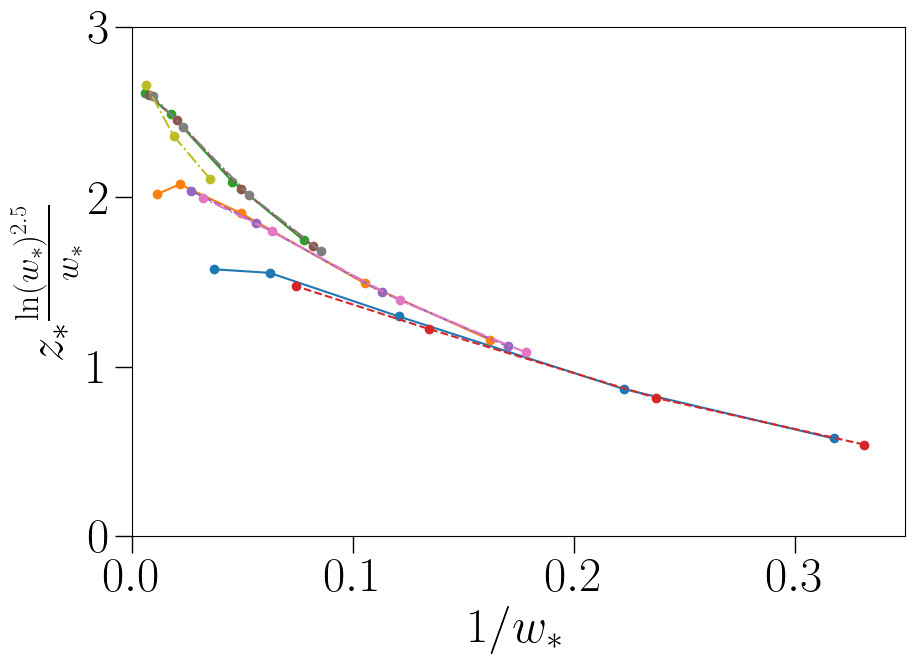}
   \caption{ Properties of the periodic solution of Eq.~(\ref{eq_ERGEdimensionless_FP_inside_periodic}) found for a range of $d$ 
   and of the prefactor $\alpha$ of the Litim/Theta regulator (given in the legend of the top panel). 
   Top: Variation with $1/w_*$ of the half-period $\varphi_*$ scaled by the anticipated factor $\sqrt{\ln w_*}$, where $w_*=\overline w(\varphi_*)$. 
   Bottom: Variation with $1/w_*$ of $z_*=\overline z(\varphi_*)$ scaled by the factor $w_*/(\ln w_*)^\mu$ with $\mu= 2.5$.
   }
    \label{fig:phiM}
  \end{center}
\end{figure}

We now discuss evidence supporting the scenario presented above. We have looked for periodic solutions of the set of equations when the 
terms in $\tilde\epsilon \varphi\partial_\varphi$ are dropped. This turns out to be an extremely hard numerical task when $w_0$ approaches 
the pole $w_P$ of the propagator. We have then lifted the condition that $\eta=2-d$ (which corresponds to $\tilde\epsilon=0$) and searched 
for solutions of the equations
\begin{equation}
\begin{aligned}
\label{eq_ERGEdimensionless_FP_inside_periodic}
 w(\varphi) = &\ \frac{2v_d}d \partial^2_\varphi \ell^{(d)}_0(w(\varphi); z(\varphi);\eta),\\
0 = &\ \eta   z(\varphi) + \beta^{(d)}_{z}(\varphi;\eta),
\end{aligned}
\end{equation}
where we have explicitly displayed the first equation only, the beta function $\beta^{(d)}_{z}$ being too long to be worth giving here (see
Appendix~\ref{appx_flows}). The equations should be solved with initial conditions $w(0)=w_0$, $w'(0)=0$,  $z(0)=1$,  $z'(0)=0$, and we 
fine-tune the value of $\eta$ to find periodic solutions $\overline w(\varphi)$ and $\overline z(\varphi)$ with the (same) half period 
$\varphi_*$. We call $\eta_{\#}(d,w_0)$ the value of $\eta$ at which this is achieved. Keeping $d$ fixed we then repeat the procedure for smaller and 
smaller values of $w_0$, thereby approaching as close as possible the pole $w_P$  and obtaining a sequence of values $\eta_\#(d,w_0)$ 
and the corresponding periodic solutions. 

We have followed this numerical procedure for the Theta/Litim regulator for which rather bulky but analytically explicit equations can be obtained. We have done it for a range of the regulator parameter $\alpha$. For each choice of $\alpha$, $d$, and $w_0$, we have then used Wolfram Mathematica to solve the system by adaptive Runge-Kutta methods of high order. We have obtained periodic solutions for a range of dimensions $0.8 \leq d \leq 1.0$ with $0.5 \leq \alpha \leq 0.8$. In Figs.~\ref{fig:chi_per} and \ref{fig:z_per} we illustrate the outcome for $w(\varphi)$, $z(\varphi)$ with $d = 0.8$, $\alpha = 0.7$, and several values of $w_0$ (in this case the pole of the propagator is in $y = 0$ and $w_P = -\alpha$). For better clarity we plot $\ln(w(\varphi) - w_P)$ and $\ln(z(\varphi))$ versus $\varphi$.

The solutions display the anticipated properties from the scenario put forward above. As $w_0$ approaches closer to the pole $w_P$ the 
half-period $\varphi_*$ at which $\overline w$ is maximum shifts to higher values while $w_*=\overline w(\varphi_*)$ rapidly increases and so does $z_*=\overline z(\varphi_*)$. (Note that $\overline z(\varphi)$ has a local minimum at the half-period $\varphi_*$, a 
local maximum being present for a value slightly less than $\varphi_*$, and of course slightly more due to the periodic nature of the solution.) 
An important feature that appears to be confirmed is the last item of the scenario in Sec.~\ref{sub_scenarioDE2}, i.e., $\varphi_*\sim \sqrt{\ln w_*}$ 
when $w_*\gg1$: see  Fig.~\ref{fig:phiM} (top). Fig.~\ref{fig:phiM} (bottom) further illustrates that $z_*$ increases more slowly than $w_*$, in a manner 
that is actually compatible with what is found in the numerical results for the full solution in the boundary layer (see Fig.~\ref{fig:zoupp+}), 
$z_*\sim w_*/[\ln w_*]^\mu$ with $\mu\approx 2.5$.

As already alluded to when discussing the LPA' case, the periodic solutions to the set of equations obtained by dropping the terms in 
$\tilde\epsilon \varphi\partial_\varphi$ display an emergent boundary layer around the half-period $\varphi_*$ when $w_0$ approaches the pole 
$w_P$. The analytic solution of this boundary layer is discussed in Appendix~\ref{sec:bl_periodic}. As it should (see Sec.~\ref{sub_scenarioDE2}), 
this boundary-layer solution and the boundary-layer solution of the full set of fixed-point equations derived in Sec.~\ref{sec:BL} coincide at leading 
order in the matching region where $1\ll \varphi\ll \varphi_*, \varphi_{\rm min}$, $1\ll w\ll w_*, w(\varphi_{\rm min})$, and 
$1\ll z\ll z_*, z(\varphi_{\rm min})$, provided  $\varphi_* \sim \varphi_{\rm min}$, $w_*\sim w(\varphi_{\rm min})$, and $z_*\sim z(\varphi_{\rm min})$.

Several additional comments are worth making. First, the numerical procedure for obtaining periodic solutions seems to be limited by the formation of boundary layers near the extrema of the functions $\overline w$ and $\overline z$ when the pole $w_P$ is approached (see Appendix~\ref{sec:bl_periodic}). We have found that the procedures used for solving the equations are limited to 16 digits of accuracy, but higher precision would be needed to resolve the solution accurately when $w_*$ and $z_*$ exceed certain values, which leads to increasingly sharper variations in the $\varphi$ dependence.

\begin{figure}[h!]
  \begin{center}
 \includegraphics[width=1\linewidth]{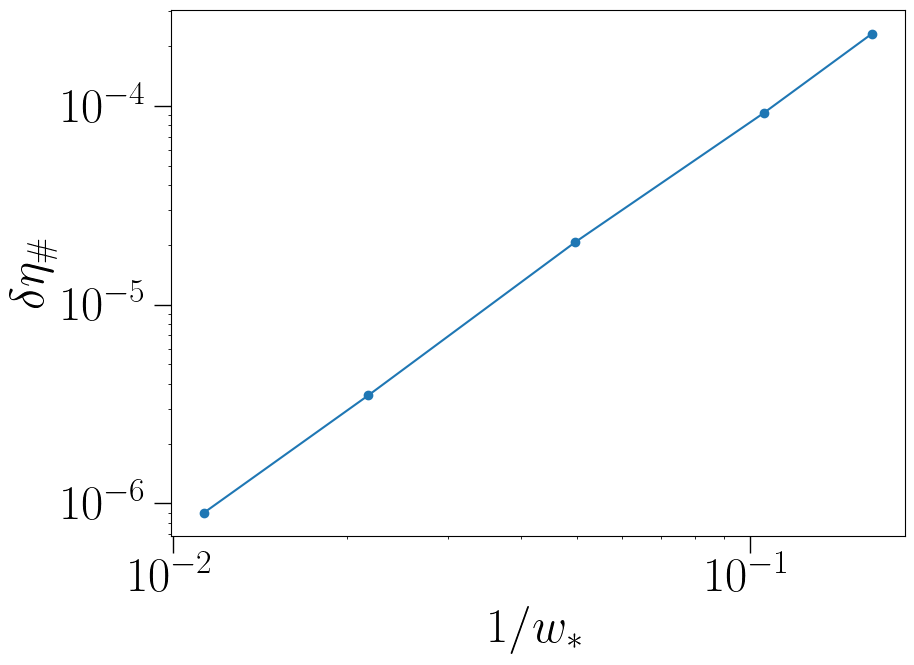}
\caption{Estimate of the variation $\delta\eta_\#$ around the periodic-solution value $\eta_\#$ for which the solution falls into either of the 
unphysical regimes described in the text when $w_*$ becomes very large (as $w_0$ approaches the pole of the propagator). Here, 
$d=0.8$ and $\alpha=0.6$. Note that this is a log-log plot.
}
    \label{fig:deltaeta}
  \end{center}
\end{figure}

Second, we can further address the motivations to focus on periodic solutions: Why 
wouldn't another kind of solution be a good candidate for the matching procedure? We have investigated this question along trajectories 
$\eta(d,w_0)$ which for a chosen $d$ (and of course $\alpha$) are either above or below the trajectory $\eta_\#(d,w_0)$ associated with periodic solutions. We have generically found that when $w_0$ approaches the pole, if $\eta$ is just slightly above $\eta_\#(d,w_0)$, then the solution 
$z(\varphi)$ diverges at some finite $\varphi$, which causes $w(\varphi)$ to diverge as well. If on the other hand $\eta$ is just slightly below 
$\eta_\#(d,w_0)$, then $z(\varphi)$ becomes negative at some finite $\varphi$, which is unphysical. The conclusion is that periodic solutions appear 
to be the only ones that make physical sense as $w_0$ approaches the pole $w_P$. Furthermore, when the value of $w_*$ is  large as 
$w_0$ approaches the pole, the variation $\delta\eta_\#$ around $\eta_\#$ that is sufficient to drive the solution in either of the above 
unphysical regimes becomes extremely small, as illustrated in Fig.~\ref{fig:deltaeta}.

\begin{figure}[h!]
  \begin{center}
 \includegraphics[width=1.0\linewidth]{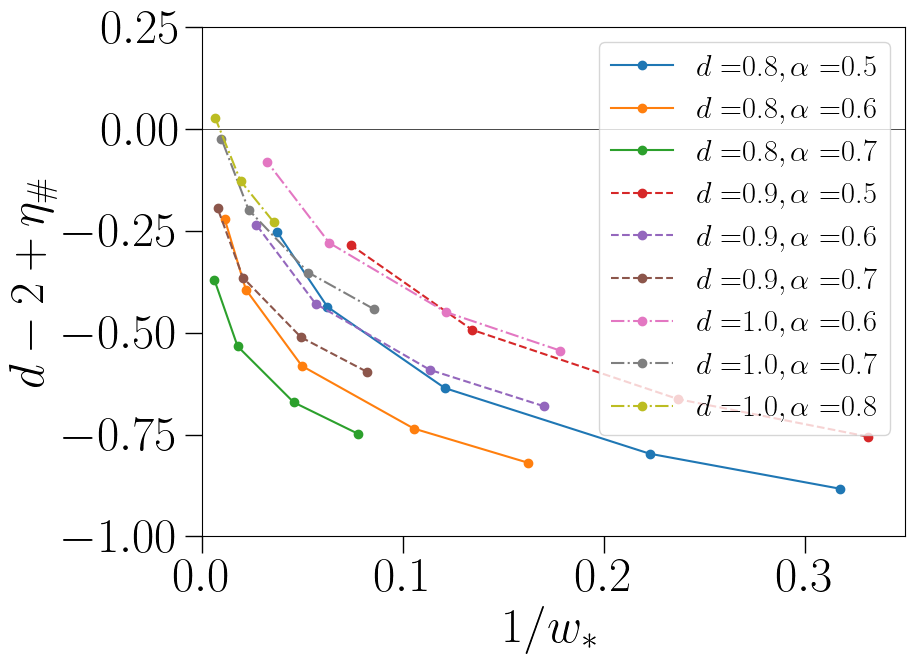}
    \caption{Variation of $d-2+\eta_\#$ with $1/w_*$ down to the lowest values of $1/w_*$ for which a periodic solution can be found. An 
    acceptable fixed-point solution at the lower critical dimension should have $d-2+\eta_\#=0$ when $1/w_*=0$. The data are obtained for a 
    range of values of $d$ and of the prefactor $\alpha$ of the Litim/Theta regulator.
    }
    \label{fig:dmtpeta}
  \end{center}
\end{figure}

All of this being said, one still needs to check that (for a given IR regulator) the periodic solution persists up to the end of the trajectory 
$\eta_\#(d,w_0)$ when $\eta_\#(d,w_0)=2-d$ and $w_0\to w_P$. This would correspond to the desired fixed-point solution that can be matched 
with the boundary-layer one (see above). We display our results in Fig.~\ref{fig:dmtpeta} where we plot $d-2+ \eta_\#$ versus $1/w_*$ for the 
already discussed periodic solutions that, as we have stressed, we are able to obtain for a limited range of $d$ and of the regulator parameter 
$\alpha$ only (we use the Litim/Theta regulator). In principle the value of the lower critical dimension $d_{\rm lc}(\alpha)$ is that for which 
$d-2+ \eta_\#=0$ when $1/w_*=0$. One can see that this seems to exclude the value $d=1$ because $d-2+ \eta_\#$ vanishes for a small yet 
nonzero $1/w_*$. The other combinations of $(d,\alpha)$ could be viable candidates but this requires extrapolating the data in some form 
or another. As we have already encountered in this problem, there could be a steep approach to zero with a diverging derivative with respect to 
$1/w_*$, e.g., where $(d-2+ \eta_\#)$ times $w_*$ goes as $\ln(w_*)$ to some power. For the Litim/Theta regulator considered here the 
data for $d=0.8$ and $0.9$ is compatible with such a behavior.
\\

\begin{figure}[h!]
  \begin{center}
\includegraphics[width=1.0\linewidth]{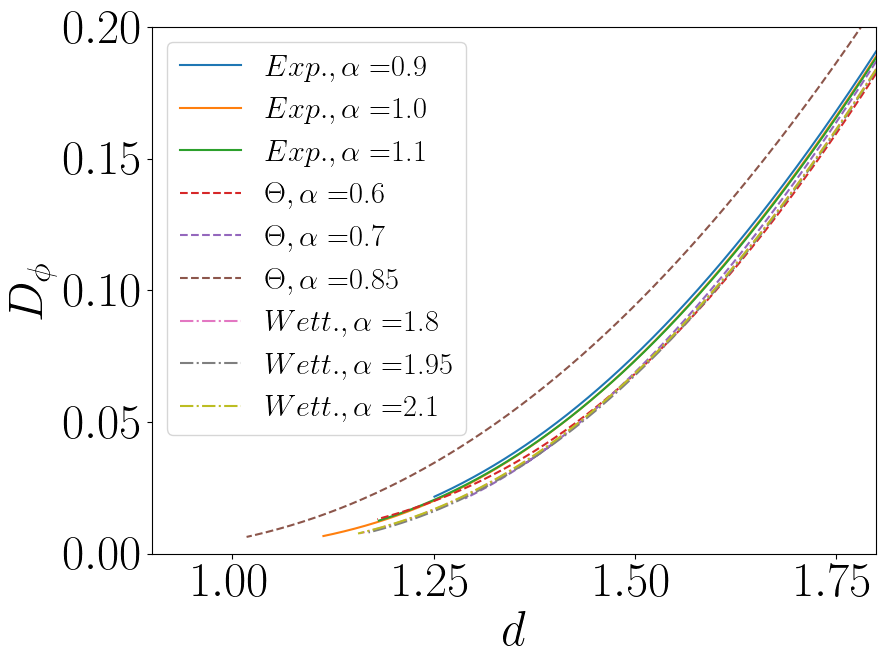}
    \caption{Scaling dimension of the field $D_\phi$ versus space dimension $d$ from the solution of the fixed-point equations at $\partial^2$ 
    order for the three IR regulators, Theta/Litim, Exponential, and Wetterich, and a range of prefactor $\alpha$. 
    }
    \label{fig_Dphi-vs-d}
  \end{center}
\end{figure}

\begin{figure}[h!]
  \begin{center}
 \includegraphics[width=1.0\linewidth]{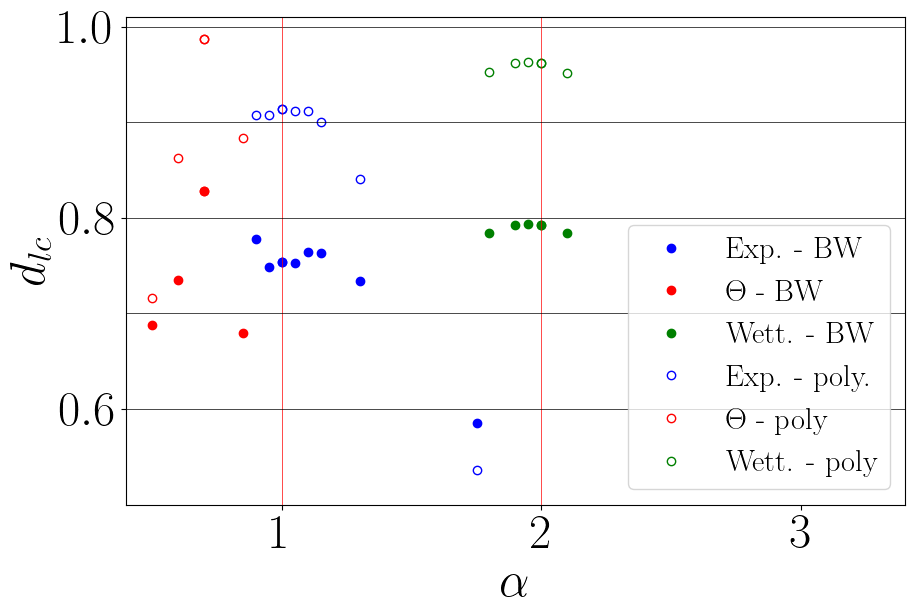}
    \caption{Estimates of the lower critical dimension $d_{\rm lc}$ from extrapolations of the curves $D_\phi$ versus $d$ shown in 
    Fig.~\ref{fig_Dphi-vs-d} for the three IR regulators, Theta/Litim (red symbols), Exponential (blue symbols), and Wetterich (green symbols). 
    The open and filled circles correspond to fits using a second-order polynomial and Eq.~(\ref{eq_BW_form}), respectively. 
    }
    \label{fig:dl_extrapol}
  \end{center}
\end{figure}

\subsection{More on the estimate of $d_{\rm lc}$ at $\partial^2$ order}
\label{sub_extrapolations}

We now go back to the numerical results obtained by numerically solving the full fixed-point equations for $d$ as low as possible. We illustrate 
the dependence of the field scaling dimension $D_\phi=(d-2+\eta)/2\approx d\tilde\epsilon$ on the dimension $d$ in Fig.~\ref{fig_Dphi-vs-d} for 
the three regulators and a range of prefactors $\alpha$. From this plot, the lower critical dimension $d_{\rm lc}$ is estimated by extrapolating 
the curves to the point where $D_\phi=0$. The most naive way to proceed is to use a (quadratric) polynomial fit to extrapolate the data and this 
gives a first set of estimates.

Before investigating an alternative formula to extrapolate the data we need to make a detour and discuss the possible connection of our 
FRG approach with the droplet theory of Bruce and Wallace.\cite{bw81_lett,bw83} As already stressed, a key result of the latter is that the 
approach to the lower critical dimension is controlled by two small parameters that are nonperturbatively related: the fractal dimension of the 
droplet surface, which goes as $d-d_{\rm lc}$ and controls the vanishing of the critical temperature $T_c$ and of the inverse of the correlation 
length exponent $1/\nu$, and the critical droplet concentration which controls the scaling dimension of the field $D_\phi=(d-2+\eta)/2$. The 
two are related by
\begin{equation}
\label{eq_BW_form}
D_\phi \propto \frac{1}{d-d_{\rm lc}}e^{-\frac{2}{d-d_{\rm lc}}},
\end{equation}
where in this exact calculation, $d_{\rm lc}=1$. Inverting the relation yields
\begin{equation}
\label{eq_BW_form2}
d-d_{\rm lc}\propto \frac 1{\ln\left(\frac 1{D_\phi}\right)},
\end{equation}
which in particular leads to $T_c$ and $1/\nu$ going to zero as  $1/\ln(\frac 1{D_\phi})$ when $D_\phi\to 0$ and $d\to 1$.

Interestingly, the solutions of the FRG fixed-point equations at the LPA' and $\partial^2$ orders also have several small parameters 
that are associated with the peculiar boundary-layer form that emerges as one approaches $d_{\rm lc}$. The natural parameter is 
$\tilde\epsilon$ which, up to a factor of $d$, is equal to $D_\phi$. Another one is $1/\ln(1/\tilde\epsilon)$ which appears in the 
characteristics of the boundary layer. As already discussed in the previous work on the LPA' approximation, the critical temperature 
$T_c$ can be related in the FRG to the field at the minimum of the effective potential, $\varphi_{\rm min}$, as\cite{lnf23}
\begin{equation}
\label{eq:Tc_dependence}
T_c\propto \frac{1}{\varphi^2_{\rm min}}, 
\end{equation}
which predicts that $T_c\propto 1/\ln(\frac{1}{\tilde{\epsilon}})\propto1/\ln(\frac 1{D_\phi})$. This is the same relation as found by 
the droplet theory.

This correspondence with the droplet theory suggests fitting the data for $D_\phi$ versus $d$ to the formula in Eq.~(\ref{eq_BW_form}). The 
outcome for $d_{\rm lc}(\alpha)$ is shown in Fig.~\ref{fig:dl_extrapol} together with the result of the fit to a second-order polynomial. The 
difference between the two procedures is of about $15\%$ and we note that we obtain values that are systematically below the exact result 
$d_{\rm lc}=1$.

\begin{figure}[h!]
  \begin{center}
 \includegraphics[width=1.0\linewidth]{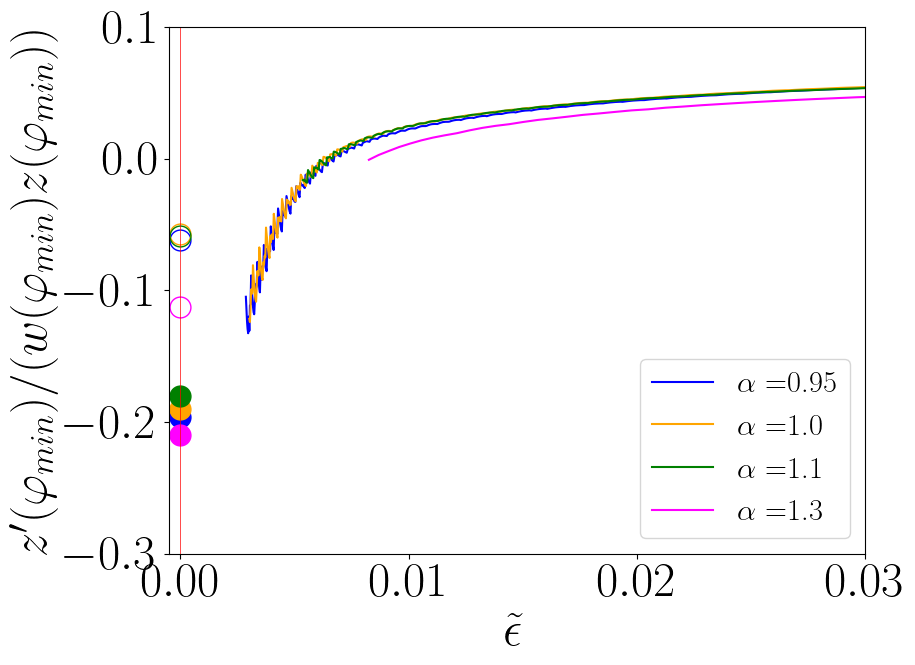}
    \caption{Ratio $z'(\varphi_{\rm min})/(w(\varphi_{\rm min})z(\varphi_{\rm min}))$ versus $\tilde\epsilon$ as obtained for small but nonzero 
    values of $\tilde\epsilon$ from the solution of the full fixed-point equations at $\partial^2$ order with the Exponential regulator and several 
    values of the prefactor $\alpha$. In $\tilde\epsilon=0$ we display the right-hand side of Eq.~(\ref{eq:zinv}) with the estimates of $d_{\rm lc}(\alpha)$ 
    found in Fig.~\ref{fig:dl_extrapol} either via a polynomial fit (open circles) or via Eq.~(\ref{eq_BW_form}) (filled circles).
    }
    \label{fig:zinvariant}
  \end{center}
\end{figure}

To further constrain the estimate of $d_{\rm lc}(\alpha)$ we use the relation that we have derived through the boundary-layer analysis of 
the fixed-point equations, Eq.~(\ref{match_de2}). Switching back from the boundary-layer functions to the original ones, one can rewrite 
Eq.~(\ref{match_de2}) as
\begin{equation}
\label{eq:zinv}
\frac{z'(\varphi_{\rm min})}{w(\varphi_{\rm min})z(\varphi_{\rm min})}=2\sqrt{\frac{2}{\pi}}-\sqrt{2}
\frac{\Gamma\left(1+\frac 1{d_{\rm lc}}\right)}{\Gamma\left(\frac 12+\frac 1{d_{\rm lc}}\right)}.
\end{equation}
The left-hand side is accessible from the fixed-point solutions at small but nonzero $\tilde\epsilon$. Extrapolating its dependence on 
the latter down to $\tilde\epsilon=0$ should give $d_{\rm lc}$ via the expression in the right-hand side. We have carried out this calculation 
for the Exponential regulator and a range of values of $\alpha$. The data are plotted in Fig.~\ref{fig:zinvariant} together with the 
expected values of the right-hand side for $d_{\rm lc}$ obtained above from fitting either a polynomial or Eq.~(\ref{eq_BW_form}). 
For the lowest $\tilde\epsilon$ that could be accessed (for $\alpha=0.95-1.1$) the data clearly seem to exclude the estimates of 
$d_{\rm lc}$ found with the polynomial fit and are more compatible with the fit through Eq.~(\ref{eq_BW_form}). However, it 
appears difficult at this point to more tightly constrain the value of $d_{\rm lc}(\alpha)$.
\\

\section{Stability analysis of the fixed point}
\label{sec:essential}

\subsection{Generalities for the $\partial^2$ approximation}
\label{sub_generalities}

As usual, the stability of the fixed point is investigated by introducing a perturbation, $w_k(\varphi)=w(\varphi) +k^\lambda \delta w(\varphi)$, $z_k(\varphi)=z(\varphi) +k^\lambda \delta z(\varphi)$, $\eta_k=\eta+k^\lambda \delta \eta$, where $w$, $z$, and $\eta$ are the fixed-point 
quantities, and linearizing the FRG flow equations, Eqs.~(\ref{eq_ERGEdimensionless}) and (\ref{eq_running_eta}), with respect to the 
perturbation. The problem turns into an eigenvalue determination with $\lambda$ corresponding to a relevant direction if it is $<0$ and 
to an irrelevant one if it is $>0$. For a more compact notation we define $ f_1(\varphi)=\delta w(\varphi)$ and $ f_2(\varphi)=\delta z(\varphi)$, 
$y_1=-(2-\eta)$ and $y_2=\eta$ and we then obtain the eigenvalue equations as
\begin{equation}
\begin{aligned}
\label{eq_linearized}
&0=\delta\eta A_i(\varphi)+ (y_i-\lambda) f_i(\varphi) +\frac{d-2+\eta}2 \varphi  f'_i(\varphi) \\&+ \sum_{j=1,2} [a_{ij}(\varphi,\eta) + 
b_{ij}(\varphi,\eta) \partial_\varphi +c_{ij}(\varphi,\eta) \partial^2_\varphi] f_i(\varphi), 
\end{aligned}
\end{equation}
where $i=1,2$, and where the functions $A_i(\varphi)$, $a_{ij}(\varphi,\eta)$, $b_{ij}(\varphi,\eta)$, and $c_{ij}(\varphi,\eta)$ are derived from 
the beta functions $\beta_w^{(d)}(\varphi;\eta)$ and $\beta_z^{(d)}(\varphi;\eta)$ with the final expressions evaluated at the fixed point; more 
details are given in Appendix~\ref{app_eigenvalue}, but the important property is that the $A_i$'s, $a_{ij}$'s, and $c_{ij}$'s are even in 
$\varphi$ while the $b_{ij}$'s are odd.

The above set of equations can be rewritten in a matrix form as
\begin{equation}
\begin{aligned}
&-\delta\eta\, \mathbf A(\varphi)= (\mathbf  y-\lambda \mathbf I) \mathbf  f(\varphi) +\frac{d-2+\eta}2 \varphi  \mathbf  f'(\varphi) \\&
+ [\mathbf  a(\varphi,\eta) + \mathbf  b(\varphi,\eta) \partial_\varphi +\mathbf  c(\varphi,\eta) \partial^2_\varphi] \mathbf  f(\varphi) 
\end{aligned}
\end{equation}
where $\mathbf  y$ is the diagonal matrix with elements $y_1$ and $y_2$ and $\mathbf I$ is the identity matrix. The matrix 
$\mathbf c$ has for elements $c_{11}=-2v_d\ell_1^{(d)}(\varphi;\eta),\, c_{12}=2v_d \partial \ell_0^{(d)}(\varphi;\eta)/\partial z(\varphi), \,
c_{21}=0, \,c_{22}=-2v_d\ell_1^{(d)}(\varphi;\eta)$, so that its determinant is equal to $[2v_d\ell_1^{(d)}(\varphi;\eta)]^2$ and is strictly 
positive, which ensures that the matrix is invertible (see Appendix~\ref{app_eigenvalue}). One can therefore recast the equation as
\begin{equation}
\begin{aligned}
\label{eq_eigenvalue_matrix}
&\Big (\mathbf I \,\partial^2_\varphi+\mathbf  c(\varphi,\eta)^{-1}[\mathbf  b(\varphi,\eta)+ \frac{d-2+\eta}2 \varphi]\partial_\varphi +
\mathbf  c(\varphi,\eta)^{-1}\times \\&[\mathbf  a(\varphi,\eta)+(\mathbf  y-\lambda \mathbf I)]\Big ) \mathbf  f(\varphi) = 
-\delta\eta \,\mathbf  c(\varphi,\eta)^{-1}\mathbf A(\varphi),
\end{aligned}
\end{equation}
which is the canonical form of a linear second-order differential equation with coefficients that are continuous over ${\rm I\!R}$. Due to 
the symmetry property of the coefficients, the eigenvectors can be sorted out into even and odd under the inversion of $\varphi$. 
Note that one expects a discrete set of eigenvalues. After fixing the initial conditions $\mathbf  f(0) = \mathbf  f_0$ and 
$\mathbf  f'(0) = \mathbf  f'_0$, there is a unique solution defined over the whole range of $\varphi$ but it exists for any 
$\lambda, \delta\eta \in {\rm I\!R}$. The selection of the possible discrete set of eigenvalues and of the associated $\delta\eta$ comes from 
considering the behavior at large field.\cite{morrisPLB94} Generically, the solution for $ f_{i=1,2}(\varphi)$ is a linear combination of a 
power law and a growing exponential [e.g., for $f_1$, $\varphi^{(\lambda+d)/(d\tilde\epsilon)}$ and 
$\exp(K_d\tilde\epsilon^2\varphi^{2+2/\tilde\epsilon})$ with $K_d$ a constant of O(1)]. Requiring that the coefficient of the exponential 
be zero in both $f_1$ and $f_2$ is then expected to restrict the allowed values of $\lambda$ to a discrete set and fix the associated 
$\delta\eta$. 

Consider first the odd sector. When evaluated in $\varphi=0$, the left-hand side of Eq.~(\ref{eq_eigenvalue_matrix}) is equal to zero while 
$\mathbf  c(\varphi=0,\eta)^{-1}\mathbf A(\varphi=0)\neq 0$. As a result one must have $\delta \eta=0$. One relevant eigenvalue is trivially 
obtained because it is associated with derivatives of the fixed-point functions: $f_1(\varphi)=K w'(\varphi)$ and $f_2(\varphi)=K z'(\varphi)$. 
It is easy to check by taking a derivative with respect to $\varphi$ of the fixed point equation that $\lambda=-(d-2+\eta)/2$ which is minus 
the scaling dimension $D_\phi$ of the field (in the effective action this is associated with a $\varphi^3$ perturbation). This eigenvalue goes 
to zero at the lower critical dimension when $\tilde\epsilon=0$.

For the even sector, we expect that there is a single relevant eigenvalue which is related to the correlation-length exponent $\nu$ by  
$\lambda=-1/\nu$. A key feature of the approach to the lower critical dimension  in the exact theory is that $1/\nu\to 0$. This is a prerequisite 
for the so-called ``essential scaling" that is observed at the lower critical dimension, $d=d_{\rm lc}$. This scaling means that, the critical 
temperature $T_c$ being equal to zero, the correlation length $\xi$ still diverges as one approaches zero temperature but in an exponential 
rather than a power-law fashion,
\begin{equation}
\xi\sim \exp(\Delta/T),
\end{equation}
where $\Delta$ is the energy scale of the localized excitations (kinks and antikinks) that proliferate and destroy the finite-temperature transition 
in $d_{\rm lc}=1$. In the droplet theory,\cite{bw81_lett,bw83} $1/\nu$ goes to zero as $d-1$ when $d\to 1$. As explained in Sec.~\ref{sub_extrapolations}, 
when translated in the small parameter $\tilde\epsilon\propto D_\phi$, this implies that $1/\nu \sim 1/\ln(1/\tilde\epsilon)$ due to the nonperturbative 
relation between $\tilde\epsilon$ and $d-d_{\rm lc}$. At the LPA' level,\cite{lnf23} we showed that $1/\nu=0$ in $\tilde\epsilon=0$ but the numerical 
results pointed to an approach in $\delta(\tilde\epsilon)\sim\tilde\epsilon\sqrt{\ln(1/\tilde\epsilon)}$ instead of $1/\ln(1/\tilde\epsilon)$. 
(In this approximation $T_c$ and $1/\nu$ appear to go to zero with qualitatively different dependences on $\tilde\epsilon$, contrary to 
the exact result.)
\\

Before discussing the eigenvalue problem within the framework of singular perturbation theory we give the numerical results obtained 
by solving the eigenvalue equations for fixed $d$ as close as possible to the lower critical dimension.
\\
 
\begin{figure}[h!]
  \begin{center}
\includegraphics[width=1.0\linewidth]{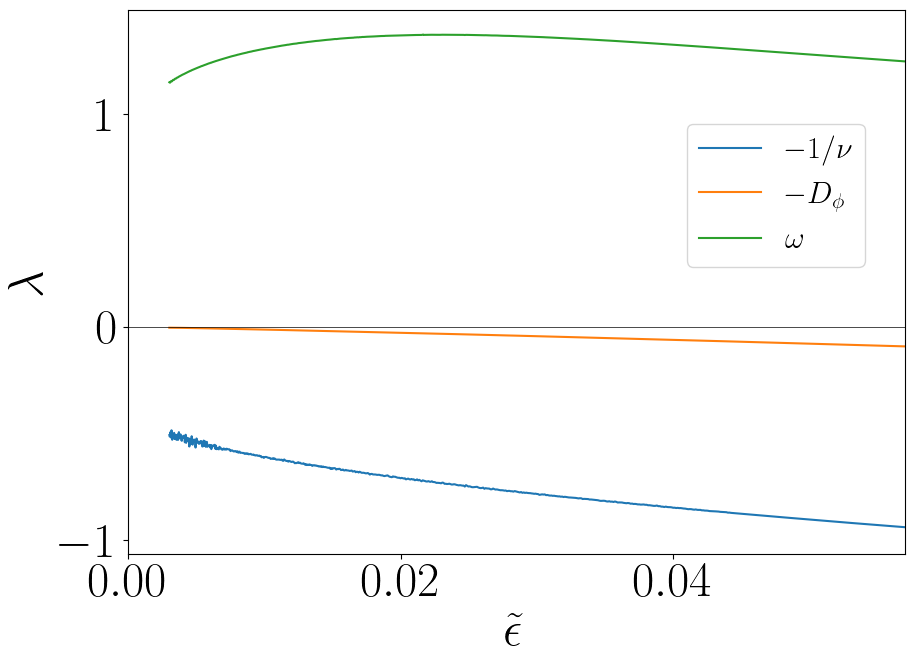}
    \caption{Variation with $\tilde\epsilon$ of the first three eigenvalues around the critical fixed point: $\lambda=-D_\phi$ in the odd sector 
    as well as $\lambda=-1/\nu$ and $\lambda=\omega$ in the even sector. We have used the Exponential regulator with prefactor $\alpha=1$. 
    }
    \label{fig_eigenvalues}
  \end{center}
\end{figure}

\begin{figure}[h!]
  \begin{center}
 \includegraphics[width=1.0\linewidth]{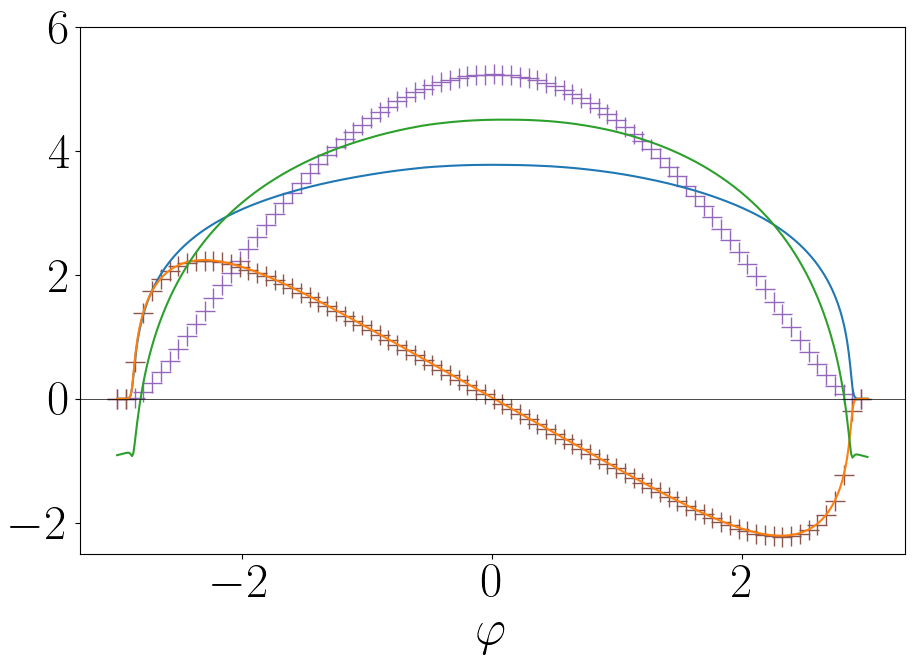}
    \caption{Eigenvectors associated with the first three eigenvalues in the direction of the function $w(\varphi)$ for $d=1.113$ where 
    $\tilde{\epsilon}=0.00301$: For improved clarity we plot $\delta w(\varphi)/[\alpha+w(\varphi)]^2$ for $\lambda=-D_\phi$ (dark red 
    symbols), $\lambda=-1/\nu$ (purple symbols) and $\lambda=\omega$ (full green line). For comparison we also plot the even function 
    $1/[\alpha+w(\varphi)]$ (full blue line) and the odd one $w'(\varphi)/[\alpha+w(\varphi)]^2$ (full orange line, which is virtually 
    indistinguishable from the eigenvector associated with $\lambda=-D_\phi$. We have used the Exponential regulator with prefactor 
    $\alpha=1.0$. 
    }
    \label{fig:3egv}
  \end{center}
\end{figure}

\subsection{Numerical results}
\label{sub_eigenvalue_numerical}

We have studied the stability of the critical fixed point as one lowers the dimension $d$. From $d=4$ down to the lowest possible $d$ for which 
we can determine the fixed point we find two and only two relevant directions around the critical fixed point: one is in the odd 
sector, and corresponds to $\lambda=-(d-2+\eta)/2=-D_\phi$, and one is in the even sector, and can be associated with 
$-1/\nu$. All of the other eigenvalues are strictly positive, i.e., irrelevant; they stay of order 1 and do not appear to go to zero as $d$ 
approaches $d_{\rm lc}$. We plot for illustration in Fig.~\ref{fig_eigenvalues} the three first eigenvalues, $-D_\phi$, $-1/\nu$, and the leading 
correction to scaling $\omega$, for the Exponential regulator with $\alpha=1$ in the vicinity of $d_{\rm lc}$ ($\tilde\epsilon\lesssim 0.5$). The 
correction-to-scaling exponent $\omega$ is close to $1$ and rather constant over the interval of $\tilde\epsilon$. The eigenvalue $-D_\phi$ 
goes to zero as $-d\tilde\epsilon$ when $\tilde\epsilon\to 0$. However, the behavior of $-1/\nu$ is more ambiguous. It decays when 
$\tilde\epsilon$ decreases but it is still of the order of $-0.5$ at the lowest accessible values of $\tilde\epsilon$. The extrapolation to zero in 
$\tilde\epsilon=0$ is therefore nontrivial, if it is indeed true that $1/\nu=0$ in this limit within the $\partial^2$ approximation. This will be 
addressed in more detail in the next subsection.

We next look at the shape of the three eigenvectors  associated with the eigenvalues $-D_\phi$, $-1/\nu$, and $\omega$. In Fig.~\ref{fig:3egv} 
we display the eigenvectors in the direction of the perturbation of the function $w$, i.e., $\delta w(\varphi)$,  for a very small value of 
$\tilde\epsilon\approx 0.003$. For a better visualization we actually plot $\delta w(\varphi)/(\alpha + w(\varphi))^2$ and we use both positive 
and negative $\varphi$ to make the symmetry of the functions clearer. We first note that all the eigenvectors appear to show a change of behavior 
near $\vert\varphi\vert \approx 2.9$, which corresponds to the location the minimum of the effective potential for this value of $\tilde\epsilon$. This 
is likely stemming from the emergence of the boundary layer already discussed at length. Next, the two eigenvectors associated with the 
relevant eigenvalues appear to go to zero when $\vert \varphi\vert$ is large, contrary to the eigenvector associated with the irrelevant eigenvalue 
$\omega$ which goes to a value of O(1). Remembering that the large-field behavior of the fixed-point function $w(\varphi)$ goes as 
$\vert\varphi\vert^{1/\tilde\epsilon}$ while that of the eigenvector $\delta w(\varphi)$ goes as $\vert\varphi\vert^{(1+\lambda/d)/\tilde\epsilon}$ 
(see above), this corresponds to the expected behavior for the relevant eigenvalues with $\lambda<0$ while for the correction-to-scaling 
eigenvalue it implies that $\omega\approx d$ [since $\delta w(\varphi)/(\alpha + w(\varphi))^2\sim \vert\varphi\vert^{(-1+\lambda/d)/\tilde\epsilon}$]. 

As anticipated, the odd eigenvector associated with the eigenvalue $\lambda=-D_\phi$ is proportional to the first derivative of the fixed-point 
function $w(\varphi)$: the two curves shown in Fig.~\ref{fig:3egv} are indistinguishable. A potentially more interesting observation which we will 
refine below is that the even eigenvector associated with $\lambda=-1/\nu$ appears very similar to the odd one when $\varphi$ is near the 
minima of the effective potential, i.e., in the boundary layer regions.

\begin{figure}[h!]
  \begin{center}
 \includegraphics[width=1.0\linewidth]{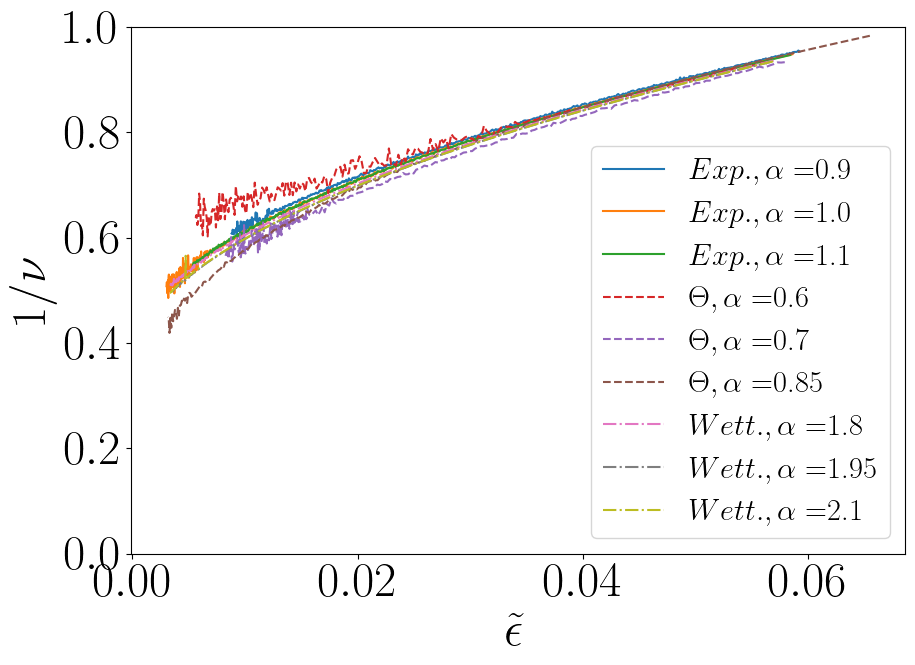}
 \includegraphics[width=1.0\linewidth]{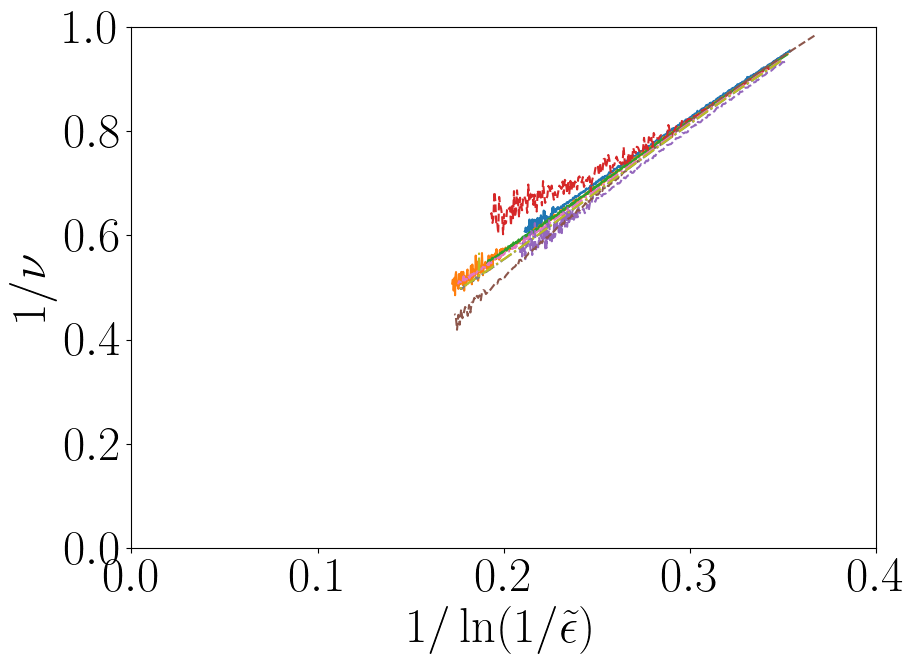}
    \caption{Variation of $1/\nu$ as one approaches the lower critical dimension for several regulator choices in the $\partial^2$ approximation. 
    Top: $1/\nu$ versus $\tilde\epsilon$. 
   Bottom: Same data plotted as $1/\nu$ versus $1/\ln(1/\tilde\epsilon)$.
   }
    \label{fig:1onu_vs_eps}
  \end{center}
\end{figure}

\subsection{Does $1/\nu\to 0$ when $\tilde\epsilon\to 0$ at $\partial^2$ order?}
\label{sub_essential-scaling?}

The main issue concerning the stability of the critical fixed point is whether $1/\nu\to 0$ as $d\to d_{\rm lc}$ and $\tilde\epsilon\to 0$. As already 
stressed this is the exact behavior with $d_{\rm lc}=1$ and it is a prerequisite for observing an essential scaling at the lower critical dimension.

In Fig.~\ref{fig:1onu_vs_eps} we display data for $1/\nu$ that is obtained for several regulators and for values of $\tilde\epsilon$ as small as 
possible. We plot the data in two different ways. In the top figure we show $1/\nu$ versus $\epsilon$. As already noticed when 
discussing Fig.~\ref{fig_eigenvalues}, it is hard to decide whether $1/\nu$ can go to zero when extrapolating to $\tilde\epsilon=0$; all curves 
do appear  to bend down when $\tilde\epsilon$ decreases but the values of $1/\nu$ reached for the lowest accessible $\tilde\epsilon$ are still 
of order 0.4-0.6. A fit to a dependence in $\delta(\tilde\epsilon)\sim \tilde\epsilon\sqrt{\ln(1/\tilde\epsilon)}$ (as numerically found satisfactory 
at the LPA' level) clearly gives a strictly positive value of $1/\nu$ in $\tilde\epsilon=0$.

In the bottom figure we next plot $1/\nu$ versus $1/\ln(1/\tilde\epsilon)$. This is motivated by the droplet theory\cite{bw81_lett,bw83} 
which predicts that 
\begin{equation}
\label{eq_1/nu_droplet}
\frac 1\nu \sim \left [\frac 1{\ln(\frac 1{\tilde\epsilon})}\right ].
\end{equation}
As seen from the figure, such a behavior is more compatible with our data, and plausible linear extrapolations to $\tilde\epsilon=0$ 
indeed give intercepts close to $1/\nu=0$. This is suggestive but by no means conclusive. We thus take an additional angle on the 
problem by studying in more detail the eigenvalue equations and the eigenvectors associated with the relevant perturbations.

\subsection{Singular perturbation analysis}
\label{sub_singular}

As done before for the fixed-point equations, we now turn to a singular perturbation analysis of the eigenvalue equations when 
$\tilde\epsilon\to 0^+$. We again expect that the eigenvectors associated with the two relevant eigenvalues can be separately 
studied in three domains --fields of O(1), large fields, and the boundary layer around the minimum $\varphi_{\rm min}$ of 
the effective potential-- with a full solution then obtained by matching the partial ones in the regions where they overlap. 

Consider the boundary layer around the minimum. We expect that the eigenvectors can be written as the fixed-point functions in 
the form
\begin{equation}
\begin{aligned}
\label{eq_eigenvectors_BL}
&\delta w(\varphi)=\frac{\delta g(x)}{\delta(\tilde\epsilon)}\\&
\delta z(\varphi)=\frac{\kappa(\tilde\epsilon)}{\delta(\tilde\epsilon)}\delta\zeta(x)
\end{aligned}
\end{equation}
with $x=(\varphi-\varphi_{\rm min})/\delta(\tilde\epsilon)$. Inserting this in the eigenvalue equations and using the same scaling analysis as 
for the fixed-point functions, we find at leading order 
 \begin{equation}
\begin{aligned}
\label{eq_eigenvalue_wBL}
0=&-(1+\frac \lambda d)\delta g(x)+\delta g'(x)+ \delta\Big [\partial^2_x\left(\frac{1}{g(x)}\right)\Big ] \\&
+C_d \frac{\delta\eta}{\tilde\epsilon}g_*'(x)+{\rm O}(\kappa(\tilde\epsilon),\delta\eta)
\end{aligned}
\end{equation}
and
 \begin{equation}
\begin{aligned}
\label{eq_eigenvalue_zBL}
0=&(\frac{(2-d)}d+\frac \lambda d)\delta \zeta(x) + \delta\zeta'(x) - \delta\Big [ \frac{\zeta''(x)}{g(x)^2}+ 4\frac{\zeta'(x)}{g(x)}\times \\&
\partial_x\left(\frac{1}{g(x)}\right) + 2\zeta(x)\left(\partial_x\left(\frac{1}{g(x)}\right)\right)^2 \Big] + C_d \frac{\delta\eta}{\tilde\epsilon}\zeta_*'(x) \\&
+{\rm O}(\kappa(\tilde\epsilon),\delta\eta)
\end{aligned}
\end{equation}
where $\delta \Big[\cdots\Big]$ denotes the linear variation due to the perturbations around the fixed-point functions, e.g., 
$\delta \Big[\partial^2_x(1/g(x))\Big]=-\partial^2_x (\delta g(x)/g_*(x)^2)$, and we have momentarily reinstalled the notation $g_*,\zeta_*$ 
and $w_*, z_*$ for the fixed point.

Note that contrary to what happens in the LPA' approximation,\cite{lnf23} $\delta\eta$ is not determined by the behavior in the boundary 
layer only, as we have chosen the conventional renormalization prescription at $\partial^2$ order that the running anomalous dimension is 
fixed by the condition in $\varphi=0$, $z_k(\varphi=0)=1$ implying $\delta z(\varphi=0)=0$. 

As shown above, for the odd eigenvector associated with $\lambda=-D_\phi\approx -d \tilde\epsilon$, one rigorously finds that $\delta\eta=0$. 
In this case, a  solution of Eqs.~(\ref{eq_eigenvalue_wBL}) and (\ref{eq_eigenvalue_zBL}) is provided by $\delta g(x)=Kg'_*(x)$, 
$\delta\zeta(x)=K\zeta'_*(x)$, with the eigenvalue $\lambda=0$ at the leading order. More generally, any solution of this kind with $\lambda=0$ 
is valid provided $\delta\eta/\tilde\epsilon\to 0$ when $\tilde\epsilon\to 0$. A possible scenario, which is borne out in the LPA' approximation 
(see below), is then that to the same partial solution in the boundary layer correspond two distinct global solutions, one odd and one even; the 
partial solutions differ in the region where $\varphi={\rm O}(1)$ while nonetheless matching with similar boundary-layer solutions in the relevant 
domain of overlap.

Unfortunately, at the $\partial^2$ level we cannot go much further with the singular perturbation treatment. Indeed, as discussed in detail in 
Sec.~\ref{sec:matching_de2}, we already do not have full analytical and numerical control of the fixed-point solutions themselves. Assuming 
that $\delta\eta/\tilde\epsilon\to 0$ and $\lambda\to 0$ when $\tilde\epsilon\to 0$, it seems plausible that the set of second-order linear equations 
obtained from Eq.~(\ref{eq_eigenvalue_matrix}) for fields of O(1),
\begin{equation}
\begin{aligned}
\label{eq_eigenvalue_matrix_O(1)}
&\Big (\mathbf I \,\partial^2_\varphi+\mathbf  c(\varphi,\eta)^{-1}\mathbf  b(\varphi,\eta)\partial_\varphi+ 
\mathbf  c(\varphi,\eta)^{-1}[\mathbf  a(\varphi,\eta)+\mathbf  y]\Big ) \mathbf  f(\varphi)\\& = 0,
\end{aligned}
\end{equation}
with initial condition $\delta z(0)=0$, has one even solution on top of the odd one (recall that we are looking for solutions able to match with the 
boundary-layer ones, which themselves should match with the power-law solutions at large field). 

If no definite conclusion comes from this treatment at the $\partial^2$ level, it is worth showing how this works at the LPA' level, where we can 
get an analytic solution.

\subsection{A detour by the singular perturbation treatment at the LPA' level}
\label{sub_SPT_LPA'}

Here we build on our previous work in [\onlinecite{lnf23}] and provide new results as well.

At the LPA', one only needs the equation for $w_k(\varphi)$ and $\eta_k$. We look for odd and even eigenperturbations with an eigenvalue equal 
to zero when $\tilde\epsilon=0$. As mentioned above this can be achieved in the boundary layer by choosing $\delta w(\varphi)=K w'(\varphi)$ 
or in a scaled form $\delta g(x)=K g'(x)$ (where $w(\varphi)$ and $g(x)$ denote the fixed-point functions). This solution matches at large field 
with the relevant power-law behavior $\delta w(\varphi)\sim \varphi^{1/\tilde\epsilon+{\rm O}(1)}$ when $\lambda=0$ and $\tilde\epsilon\to 0^+$. 
This is possible because one can also show that within the boundary layer $\delta\eta=0$ at leading order; 
see Appendix~\ref{app_eigenvalue-LPA'}.

The odd or even property of the eigenvector under $Z_2$ field inversion is determined by the solution around $\varphi=0$ and for fields 
of O(1). The eigenvalue equation at leading order when $\tilde\epsilon=0$ (with $\lambda=0$) reads
\begin{equation}
0= -d\,\delta w(\varphi)-2v_d \frac{\partial^2}{\partial \varphi^2}[\ell^{(d)}_1(\overline w(\varphi))\delta w(\varphi)],
\end{equation}
with $\ell^{(d)}_1(w)=-\partial_w\ell^{(d)}_0(w)$, while the corresponding fixed-point equation satisfied by $\overline w(\varphi)$ is
\begin{equation}
0= -d \overline w(\varphi)+2v_d \frac{\partial^2}{\partial \varphi^2}[\ell^{(d)}_0(\overline w(\varphi))],
\end{equation}
where the overline denotes the periodic fixed-point solution for $\varphi={\rm O}(1)$ and $\tilde\epsilon\to0$. An odd eigenvector is obviously 
given by $\delta w(\varphi)=K \overline w'_*(\varphi)$ and it matches with the solution in the boundary layer. It is easy to show that there is also 
an even eigenvector of the form $\delta w(\varphi)=c(\varphi) \overline w'_*(\varphi)$ with
\begin{equation}
\begin{aligned}
\label{eq:c_limits}
c(\varphi) &\sim \frac{c_0}{\varphi}, \;\;{\rm if }\; \varphi\to 0 \\&
   \sim  c_1,  \;\;{\rm if }\;  1\ll\varphi\ll \varphi_*,
\end{aligned}
\end{equation}
with $c_0$ and $c_1$ two constants, and where $\varphi_*$ is the half-period of the fixed-point solutions (we only consider the 
range $\varphi\geq 0$ due to the symmetry). More details are given in Appendix~\ref{app_eigenvalue-LPA'}. This solution matches 
with the boundary-layer one when $1\ll\varphi\ll \varphi_*$ (provided the fixed-point solution $\overline w(\varphi)$ itself matches 
with the fixed-point solution in the boundary layer, which imposes a condition on $w(\varphi=0)$\cite{lnf23}).

The singular perturbation treatment thus provides us with one odd and one even eigenvector associated with the eigenvalue $\lambda=0$ 
when $\tilde\epsilon\to 0$, with both eigenvectors being proportional to the first derivative of the fixed-point function in the 
boundary layer around the minimum of the potential. This is this latter property that we will now check in the case of the 
$\partial^2$ level of approximation.

\begin{figure}[h!]
  \begin{center}
 \includegraphics[width=1.0\linewidth]{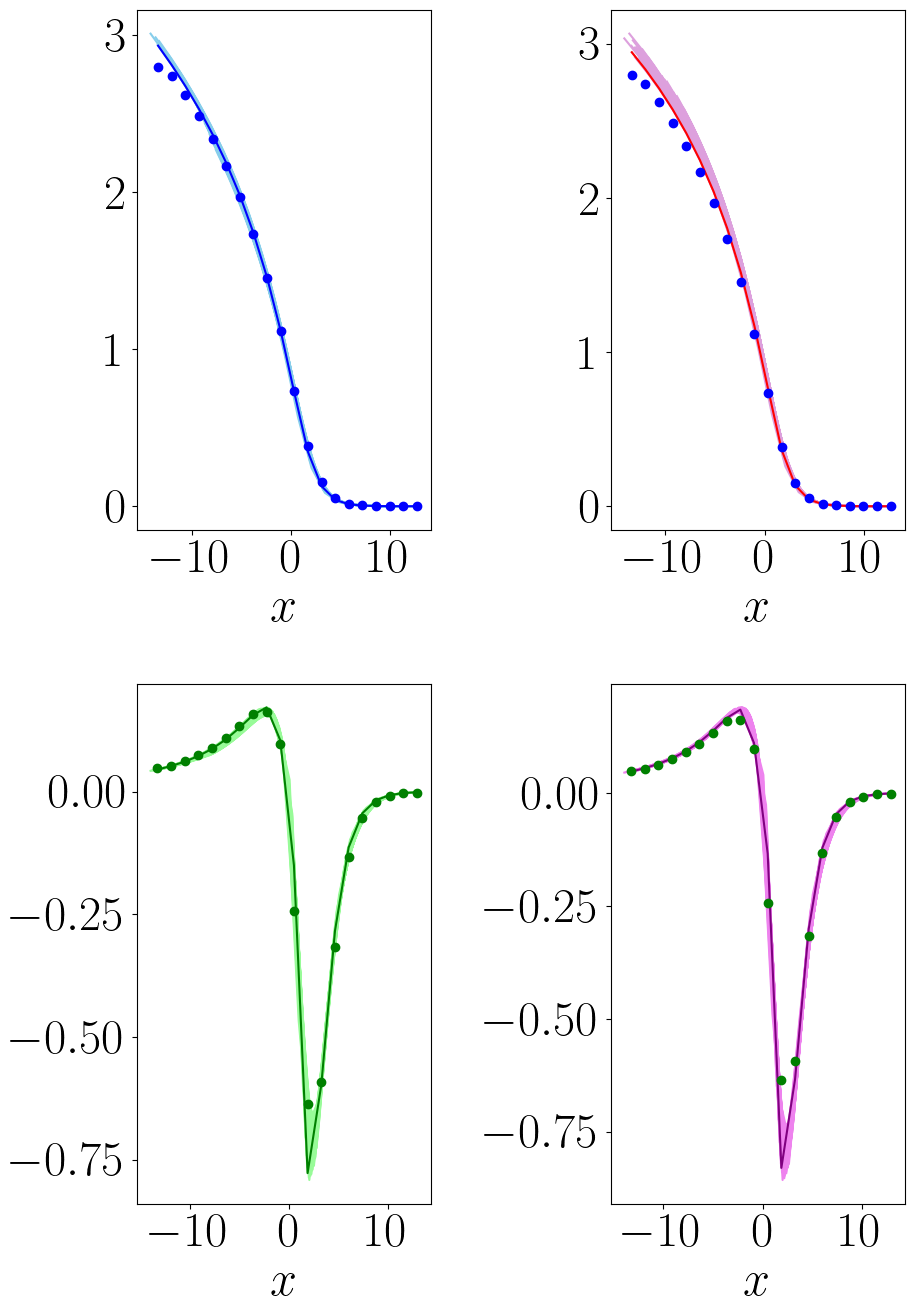}
    \caption{Zoom in onto the eigenvectors associated with $\lambda=-1/\nu$ (left) and $\lambda=-D_\phi$ (right) in the region near the 
    minimum of the effective potential. The functions $\delta g(x)/g(x)^2=(1/\delta(\tilde\epsilon))\delta w(\varphi)/w(\varphi)^2$ (top panels) 
    and $\delta \zeta(x)=(\delta(\tilde\epsilon)/\kappa(\tilde\epsilon)) \delta z(\varphi)$ (bottom panels) are plotted in terms of the boundary-layer 
    variable $x=(\varphi-\varphi_{\rm min})/\delta(\tilde\epsilon)$. The full lines are the functions obtained for $d=1.113$ (which corresponds 
    to $\tilde{\epsilon}=0.00301$) and the lightly colored lines represent all the eigenvectors between $d=1.2$ ($\tilde{\epsilon}=0.006$) and 
    $d=1.113$ collapsed with the boundary-layer ansatz in Eq.~(\ref{eq_eigenvectors_BL}). We also plot the first derivatives 
    of the fixed-point functions, $w'(\varphi)/w(\varphi)^2$ divided by $\delta(\tilde\epsilon)$ (symbols in the top panels) and 
    $z'(\varphi)$ dicvided by $\kappa(\tilde\epsilon)/\delta(\tilde\epsilon)$ (symbols in the bottom panels). The superimposition with 
    the eigenvectors is done with an arbitrary proportionality constant which is different for the two eigenvalues but which is the same 
    for the $\delta w$ and the $\delta z$ directions. 
    In these plots we have used the Exponential regulator with a prefactor $\alpha=1$.}
    \label{fig:BL_eigenvectors_de2}
  \end{center}
\end{figure}

\subsection{Further numerical evidence for $1/\nu=0$ in $\tilde\epsilon=0$ at $\partial^2$ order}

We now go back to our numerical study of the eigenvectors at small but nonzero $\tilde\epsilon= 0$. We focus on the two eigenvectors 
associated with the relevant directions in the vicinity of the minimum of the effective potential, i.e., we zoom in onto the region near 
the minima for plots similar to Fig.~\ref{fig:3egv}. We do this for both $\delta w(\varphi)$ and $\delta z(\varphi)$ which we rescale with 
the form in Eq.~(\ref{eq_eigenvectors_BL}). We display in Fig.~\ref{fig:BL_eigenvectors_de2} $\delta g(x)/(\alpha+ w(\varphi))^2$, which 
provides a better visualization as we argued before,  and $\delta \zeta(x)$, with $x=(\varphi-\varphi_{\rm min})/\delta(\tilde\epsilon)$, for 
both $\lambda=-1/\nu$ (even eigenvector) and $\lambda=-D_\phi$ (odd eigenvector). We first observe that the collapse onto the 
boundary-layer scaling curves is very good for the range of $\tilde\epsilon$ from $0.006$ ($d=1.2$) down to the lowest attainable value 
$0.003$ ($d\approx 1.113$). This strongly supports that this boundary-layer scaling persists down to $\tilde\epsilon= 0$.

Strikingly, we also find that the eigenvectors for $-1/\nu$ and $-D_\phi$ are virtually superimposable up to a constant proportionality factor 
which is the same for the perturbation in the $w$ direction and in the $z$ direction. Furthermore, the eigenvectors are extremely close to the 
derivative of the associated fixed-point function. As the convergence of the eigenvectors to their form when $\tilde\epsilon\to 0$ appears 
fast, much faster than the convergence of, say, $1/\nu$ (see Fig.~\ref{fig:1onu_vs_eps}), this then gives solid evidence that the asymptotic 
forms of the eigenvectors for both  $-1/\nu$ and $-D_\phi$ in the boundary layer are proportional to the derivatives of the fixed-point functions.  
As discussed in Sec.~\ref{sub_essential-scaling?}, this implies that the eigenvalues are zero when $\tilde\epsilon\to 0$. 

This is exactly what is found in the LPA' approximation and it corroborates that the LPA' scenario is also valid at the $\partial^2$ level, 
then supporting that $1/\nu=0$ in $\tilde\epsilon= 0$. 
\\

\section{Conclusion}
\label{sec:conclusion}

In this paper we have pursued our investigation of how the generic nonperturbative approximation scheme of the functional RG that takes the form of a 
truncated expansion of the running effective-action functional in derivatives of the field around a uniform background is able to describe 
physical situations in which the long-distance properties are controlled by strongly nonuniform renormalized configurations such as 
instantons and droplets. 

We have focused on the approach to the lower critical dimension $d_{\rm lc}$ in the $Z_2$-symmetric scalar $\varphi^4$ theory (Ising 
universality class) and pushed our previous related investigation\cite{lnf23} to the second order of the derivative expansion. The numerical and 
the analytical analyses require solving two coupled differential equations for the functions of the field characterizing the critical fixed point 
and both turn out to be much more involved than at the lower order. We are therefore unable to obtain a full control of the numerics (although 
we can solve the fixed point equations down to relative distance of $10^{-3}$ of $d_{\rm lc}$) and of the analytical treatment (although we build 
on the singular perturbation approach previously developed\cite{lnf23}). We nonetheless provide strong evidence that, as at the lower 
order, the convergence of the fixed-point functions in the limit where $d\to d_{\rm lc}$ is nonuniform in the field and involves the emergence 
of a boundary layer of shrinking width near the minima of the effective potential.

As already stressed in the Introduction, nothing guarantees that the approximate NPFRG treatment recovers the exact result $d_{\rm lc}=1$. 
The value of $d_{\rm lc}$ is indeed found to depend on the specific form of the IR regulator used to progressively introduce the fluctuations 
of the field, and for the 3 different forms that we have considered we obtain an optimal value around $0.8-0.9$, i.e., below the exact result. 
This explains why NPFRG analyses directly studying $d=1$ are unable to describe the expected features of the lower critical dimension of 
Ising-like models, in contrast to similar studies of the O($N$) model where the lower critical dimension of $2$ and the Goldstone modes are 
easily retrieved.\cite{berges02,dupuis21} A prospect of future NPFRG studies would be to investigate whether one can find a specific IR 
regulator that predicts $d_{\rm lc}=1$ at the second-order of the derivative expansion, which would for instance allow one to study all regimes 
of quantum barrier tunneling or of the Schmidt transition in a resistively shunted Josephson junction. Another interesting question 
for the NPFRG concerns the mechanism by which the critical fixed point disappears at the lower critical dimension. The most likely scenario 
is the collapse with the zero-temperature fixed point associated with the ordered (symmetry-broken) phase. The latter has already been 
studied within the derivative expansion,\cite{berges02,dupuis21,pelaez-wschebor2016} but only in dimension $d=2$ and above. Extending the 
investigation below $d=2$ and toward the putative lower critical dimension might bring some new insight into the problem.

Interestingly, the nonuniform convergence and the boundary layer found in the NPFRG treatment provide a mechanism for generating 
two different small parameters that are nonperturbatively related as $d\to d_{\rm lc}$: one, denoted here by $\tilde\epsilon$, which controls 
the dimension of the field $D_\phi$ as it goes to zero and another, $1/\ln(1/\tilde\epsilon)$, which controls the vanishing of the critical 
temperature $T_c$ and, plausibly but not fully established, of the inverse correlation length $1/\nu$. (It is worth stressing 
that the latter result of the second order of the derivative expansion was not found at the lower LPA' approximation previously 
studied.\cite{lnf23}) This is precisely what is obtained in the RG theory of Bruce and Wallace which directly focuses on the droplet configurations 
that dominate the critical behavior in low dimensions: $\tilde\epsilon$ can be put in correspondence with the critical droplet concentration and 
$1/\ln(1/\tilde\epsilon)$ with the fractal dimension of the droplet surface, although no explicit description in terms of droplets is introduced 
in the truncated derivative expansion of the NPFRG. In any case this confirms the ability of the (approximate) NPFRG to capture, at least 
partially, the influence of strongly nonuniform field configurations and opens the possibility to study other (more interesting) problems where such configurations are important, such as in systems with quenched disorder.

\acknowledgements
We thank Adam Ran\c con for numerous discussions related to this topic over the past years. We acknowledge the support of the Croatian 
Scientific Foundation grant HRZZ-IP-2022-9423, ``Cogito"- Partnership Hubert Curien bilateral project with France (LNF and IB) as well as the 
QuantiXLie Centre of Excellence (LNF and IB), a project cofinanced by the Croatian Government and European Union through the European 
Regional Development Fund - the Competitiveness and Cohesion Operational Programme (Grant KK.01.1.1.01.0004).

\newpage

\appendix

\section{NPFRG flow equations at $\partial^2$ order}
\label{appx_flows}

The beta function describing the NPFRG flow of the dimensionless effective potential $u_k(\varphi)$ at $\partial^2$ order is given 
by\cite{berges02}
\begin{equation}
\begin{aligned}
\label{eq_beta_potential}
\beta_u^{(d)}(\varphi;\eta)=2 v_d \ell_0^{(d)}(u''(\varphi);z(\varphi);\eta),
\end{aligned}
\end{equation}
where $v_d^{-1}=2^{d+1} \pi^{d/2}\Gamma(d/2)$; $\ell_n^{(d)}$ is a (strictly positive) dimensionless threshold function defined by
\begin{equation}
\begin{aligned}
\label{eq_threshold_ell}
&\ell_n^{(d)}(w;z;\eta)= \\&-\left (\frac{n+\delta_{n,0}}2\right )\int_0^{\infty}dy y^{\frac d2}\frac{\eta r(y)+2yr'(y)}{(y[z+r(y)]+w)^{n+1}}
\end{aligned}
\end{equation}
where the dimensionless infrared cutoff function (or IR regulator) $r(y)$ is obtained from the dimensionful one, $R_k(q^2)$, 
which is introduced in Eq.~(\ref{eq_regulator}), through 
\begin{equation}
\label{eq_IRregulator}
R_k(q^2)=Z_k k^2 y\, r(y) \;\;\; {\rm with}\;\; y=\frac {q^2}{k^2}
\end{equation}
with $k$ the running IR cutoff and $Z_k$ the dimensionful field renormalization (from which the running anomalous dimension is defined 
by $\eta_k=-k\partial_k \ln Z_k$).

From the exact FRG equation for the 2-point 1-PI correlation function evaluated for a uniform field configuration one can extract the 
beta functional for the dimensionless field renormalization function $z_k(\varphi)$, which when truncated at $\partial^2$ order 
reads\cite{berges02}
\begin{equation}
\begin{aligned}
\label{eq_beta_fieldrenormalization_1}
&\beta_z^{(d)}(\varphi;\eta)= - \frac{4v_d}d u'''(\varphi)^2m_{4,0}^{(d)}(u''(\varphi);z(\varphi);\eta) - \\&
\frac {8v_d}d u'''(\varphi)z'(\varphi)m_{4,0}^{(d+2)}(u''(\varphi);z(\varphi);\eta)\\&
- \frac{4v_d}d z'(\varphi)^2 m_{4,0}^{(d+4)}(u''(\varphi);z(\varphi);\eta) - 2v_d z''(\varphi) \times \\&
\ell_1^{(d)}(u''(\varphi);z(\varphi);\eta)+  4 v_d u'''(\varphi) z'(\varphi) \ell_2^{(d)}(u''(\varphi);z(\varphi);\eta) \\&
+ 2 v_d \frac{1+2d}d z'(\varphi)^2 \ell_2^{(d+2)}(u''(\varphi);z(\varphi);\eta),
\end{aligned}
\end{equation}
where $m_{n,0}^{(d)}$ is another (strictly positive) threshold function defined as
\begin{equation}
\begin{aligned}
\label{eq_threshold_m}
&m_{n,0}^{(d)}(w;z;\eta)= \frac 12 \int_0^{\infty}dy y^{\frac d2} \frac{z+(yr(y))'}{(y[z+r(y)]+w)^n}
\bigg [2\eta (yr(y))'\\&
+4(y^2r'(y))' -n \,\frac {y[z+(yr(y))'][\eta r(y)+2yr'(y)]}{y[z+r(y)]+w}  \bigg ].
\end{aligned}
\end{equation}
To derive Eq.~(\ref{eq_beta_fieldrenormalization_1}) we have neglected the higher-order terms in Eq.~(\ref{eq_DE}) which involve four 
spatial derivatives and more. The present $\partial^2$ approximation is fully characterized by the two functions $U_k(\phi)$ and $Z_k(\phi)$, 
or their dimensionless counterparts, $u_k(\phi)$ and $z_k(\phi)$. (Three more functions are required at the order O($\partial^4$), etc.)
\\

\section{Alternative scalings for the field renormalization function in the boundary layer}
\label{app:marginal_cases}

We briefly discuss alternatives for the scaling of the field renormalization function $z(\varphi)$ in the boundary layer, in which it scales differently 
than $\kappa(\tilde{\epsilon})/\delta(\tilde{\epsilon})$ with $\kappa(\tilde{\epsilon})\to 0$ when $\tilde{\epsilon}\to 0+$. 

\subsection{$z={\rm O}(1)$ in the boundary layer}

If  $z(\varphi)\equiv z(x)={\rm O}(1)$ in the boundary layer, then the corresponding fixed-point equation becomes at leading order
\begin{equation}
\begin{aligned}
\label{canonical_zO1_eq}
 & 0= (2-d)z(x) +d \frac{g'(0)}{g(0)^3}z'(x)-d\Big[ \frac{z''(x)}{g(x)^2} + 4\frac{z'(x)}{g(x)} \times \\&
  \partial_x\left(\frac{1}{g(x)}\right) +\left(2z(x)-\alpha^*(d)\frac{3d-2}{2d}\right)\left(\partial_x\left(\frac{1}{g(x)}\right)\right)^2 \Big],
 \end{aligned}
\end{equation}
where $\alpha^*(d)$ is a regulator dependent function of $d$, e.g., for the Theta or Litim regulator $\alpha^*=\alpha$ and for the Exponential 
regulator $\alpha^*=2^{-(1+\frac{d}{2})}\alpha$. 

In this case one easily shows that, when $x\to -\infty$, $z(x)$ asymptotically goes as
\begin{equation}
  \label{match_z}
  z(x)=z_0+{\rm O}\left(\frac1{\ln(\vert x\vert)}\right),
\end{equation}
with 
\begin{equation}
\label{plateau_z}
  z_0=\alpha^*(d)\frac{3d_l-2}{4 d_l} .
\end{equation}

We next have to consider the matching with the solution of the fixed-point equations for $\varphi$ of O(1) (then dropping the terms in 
$\tilde\epsilon\varphi\partial_\varphi$ and neglecting terms of order $\tilde\epsilon$). We solve the latter numerically for $\varphi$ between 
$0$ and a maximum field $\varphi_M$ by using the Runge-Kutta method implemented in Wolfram Mathematica. For matching with the 
boundary-layer solution when $\varphi\gg 1$ yet $x\to -\infty$,  we take the boundary conditions in a field $\varphi_{M}\gg 1$ satisfying 
$\varphi_{\rm min}-\varphi_{M}\sim (\tilde\epsilon\varphi_{\rm min})^a$ with $0<a<1$ and $\varphi_{\rm min}\sim \sqrt{\ln(1/\tilde\epsilon)}$ 
(see Secs.~\ref{sec:BL} and \ref{sec:matching_de2}). Then we choose 
$w(\varphi_{M})\sim \tilde\epsilon^{-a}[\ln(1/\tilde\epsilon)]^{-(1+a)/2} \gg 1$ [see Eqs.~(\ref{eq_g(x)_matching}) and (\ref{bl_ansatz_w})] 
and $z(\varphi_{M})=z_0$ where $z_0$ is given by Eq.~(\ref{plateau_z}).

We find a unique, stable, solution such that $w(\varphi)\to 0$ 
and $z(\varphi)\to 0$ as $\varphi\to 0$, which clearly cannot correspond to a physical fixed-point solution at criticality.

\subsection{$z\propto \frac{1}{\delta(\tilde{\epsilon})}$ in the boundary layer}

If $z(\varphi)\approx \zeta(x)/\delta(\tilde{\epsilon})$ in the boundary layer, i.e., $\kappa(\tilde\epsilon)={\rm O}(1)$, the fixed-point equations 
drastically change because the propagator in the boundary layer now takes the form
\begin{equation}
p(y;w(\varphi),z(\varphi))\approx\frac{\delta(\tilde{\epsilon})}{g(x)+ y[r(y)+\zeta(x)]},
\end{equation} 
with $\zeta$ and $g$ strictly positive functions of O(1). An important consequence is that the equation for $g(x)$ no longer decouples from 
that of $\zeta(x)$. Furthermore, the integration over the internal momentum cannot be performed in closed form and the equations remain 
dependent on the IR regulator. 

We do not wish to go into the detail of the corresponding calculations and we just give the main results. 

We find that for $\varphi_M$ in the matching region where $w\gg 1$ and $x\to -\infty$, $w$ and $z$ should behave as 
\begin{eqnarray}
\label{match_large_w}
w(\varphi_{M})\sim (\varphi_{\rm min}-\varphi_{M})^{-\tau_w}\\
\label{match_large_z}
z(\varphi_{M})\sim (\varphi_{\rm min}-\varphi_{M})^{-\tau_z},
\end{eqnarray}
and the boundary-layer equations are then dominated by the nontrivial part of the beta functions only if
\begin{equation}
\label{eq_tau}
\tau_z>\frac{4-(4-d)\tau_w}{d},
\end{equation}
in which case the dimensional parts are subleading. This requirement seems more natural than having the fixed-point equations dominated by their trivial dimensional part.

We then solved numerically   for $\varphi$ between $0$ and $\varphi_{M}$ the equations derived for fields of O(1) by neglecting terms in 
$\tilde\epsilon\varphi\partial_\varphi$. With boundary conditions in $\varphi_M$ of the form given by Eqs.~(\ref{match_large_w},
\ref{match_large_z}),\ref{eq_tau}) and with a range of regulator prefactors $\alpha={\rm O}(1)$, no physical fixed-point solution was found, 
whatever the choice of $\tau_w,\tau_z$ (obeying Eq. (\ref{eq_tau})). Indeed, $z(\varphi\to 0)$ always becomes very large, in contradiction with the condition that 
$z(\varphi=0)=1$.
\\

\section{Boundary-layer periodic solutions}
\label{sec:bl_periodic}

We note that the equations derived for fields of O(1) when dropping the terms in $\tilde\epsilon\varphi\partial_\varphi$, i.e., formally setting $\tilde\epsilon=0$ in the equations, also admit boundary-layer solutions around the minimum $\varphi_*$ of the periodic solution $\overline w(\varphi)$. 
We call them ``boundary-layer periodic" solutions. Besides the fact that they may be interesting when considering the matching procedure between 
the periodic solution $(\overline w(\varphi),\overline z(\varphi))$ of the $\tilde\epsilon=0$ equations and the (physical) boundary-layer fixed-point equations, they may also be relevant in the context of the Sine Gordon model.\cite{daviet23}\footnote{Private communication with Romain Daviet}

Denoting the width of the "periodic boundary layer" by $\overline\delta$ and assuming that
\begin{equation}
\begin{aligned}
&\overline{w}(\varphi)=\frac{\overline{g}(\overline x)}{\overline\delta}\\&
\overline{z}(\varphi)=\frac{\overline \kappa}{\overline\delta}\overline{\zeta}(\overline x),
\end{aligned}
\end{equation}
with $\overline x=(\varphi-\varphi_*)/\overline\delta$, we can repeat the derivation of the boundary-layer fixed point equations but starting from 
the equations with $\tilde{\epsilon}=0$. This gives at leading order
\begin{equation}
\begin{aligned}
\label{eq:blLAM_fpe_gz}
&0=-\overline{g}(\overline{x})+\partial^2_{\overline{x}}\left(\frac{1}{\overline{g}(\overline{x})}\right)\\&
%\label{eq:blLAM_fpe_z}
0=\frac{(2-d)}d \overline{\zeta}(\overline{x})-\Big[\frac{\overline{\zeta}''(\overline{x})}{\overline{g}(x)^2}+ 4\frac{\overline{\zeta}'(x)}{\overline{g}(\overline{x})}\partial_{\overline{x}}\left(\frac{1}{\overline{g}(\overline{x})}\right)\\&
+2\overline{\zeta}(\overline{x})\left(\partial_{\overline{x}}\left(\frac{1}{\overline{g}(x)}\right)\right)^2\Big ].
\end{aligned}
\end{equation}
The only difference with the physical boundary-layer equations is the absence of the term in $\zeta'(x)$ in the second equation: compare with Eq.~(\ref{eq_g-zeta_BL}). 

Similarly to what is done in Sec.~\ref{sec:BL}, we proceed with a change of variable,
\begin{equation}
\label{eq:aux_W}
\overline X(\overline{x})=\partial_{\overline{x}}\left(\frac{1}{\overline{g}(\overline{x})}\right),
\end{equation}
and we introduce the functions
\begin{eqnarray}
\label{eq:defaLAM}
\overline{a}(\overline X)\equiv\frac{1}{\overline{g}(\overline{x})}\\
\label{eq:defbLAM}
\overline{b}(\overline X)\equiv\frac{\overline{\zeta}(\overline{x})}{\overline{g}(\overline{x})^2},
\end{eqnarray}
so that the periodic boundary-layer equations read
\begin{eqnarray}    
\label{eq:blLAM_fpe_a}
& \overline{a}'(\overline X)-\overline X\, \overline{a}(\overline X)=0\\
   \label{eq:blLAM_fpe_b}
&\overline{b}''(\overline X)-\overline X\, \overline{b}'(\overline X) - \frac{2+d}{d} \,\overline{b}(\overline X)=0.
\end{eqnarray}

Because they correspond to the periodic solution, the solution of the above equations should be symmetric in the inversion 
$\overline X\to -\overline X$ and we find them as
\begin{eqnarray}
\overline{a}(\overline X)=\overline{a}_0 e^{\frac{\overline X^2}{2}}\\
\overline{b}(\overline X)= {}_1F_1\left(\frac{q}{2},\frac{1}{2},\frac{\overline X^2}{2}\right)\,\overline{b}_0,
\end{eqnarray} 
with $\overline{a}_0$ and $\overline{b}_0$ some constants and $q=(2+d)/d$as for the physical boundary-layer solution. 

We can now study the matching of the periodic and physical boundary-layer solutions in the matching region where $\overline x\to -\infty$ 
(for the former) and $x\to -\infty$ (for the latter). Then, $\overline X, X$ go to $-\infty$ as $-\sqrt{2\ln(\vert\overline x\vert)}$ and 
$-\sqrt{2\ln(\vert x\vert)}$ respectively. The leading orders correspond provided one chooses $\overline{a}_0=\sqrt{2\pi}$ and 
$\overline{b}_0=\hat b\lim_{X\to-\infty}\frac{H_{-q}(\frac{X}{\sqrt{2}})}{{}_1F_1(\frac{q}{2},\frac{1}{2},\frac{X^2}{2})}$, where we recall 
that $\hat b=2^{q-1}\sqrt{\pi}\Gamma((1+q)/2)$ (see Sec.~\ref{sec:BL}); this applies for any value of $q$.
\\

\section{System of eigenvalue equations at $\partial^2$ order}
\label{app_eigenvalue}

To obtain the eigenvalue equations we start from the dimensionless flow equations of $w_k(\varphi)$ and $z_k(\varphi)$,
\begin{equation}
\begin{aligned}
%\label{eq_ERGEdimensionless}
\partial_t w_k(\varphi)\vert_\varphi =&-(2-\eta_k) w_k(\varphi) +\frac{(d-2+\eta_k)}{2}\varphi w_k'(\varphi) \\&+ \beta^{(d)}_{w}(\varphi;\eta_k)
\end{aligned}
\end{equation}
and
\begin{equation}
\begin{aligned}
\partial_t z_k(\varphi)\vert_\varphi = \eta_k z_k(\varphi) +\frac{(d-2+\eta_k)}{2}\varphi z_k'(\varphi) + \beta^{(d)}_{z}(\varphi;\eta_k),
\end{aligned}
\end{equation}
where from Appendix~\ref{appx_flows},
\begin{equation}
\begin{aligned}
\label{eq_beta_w_DE2}
&\beta_w^{(d)}(\varphi;\eta)\equiv \beta_w^{(d)}(w(\varphi),w'(\varphi),w''(\varphi);z(\varphi),z'(\varphi),z''(\varphi);\eta)\\&
=2 v_d \partial_\varphi^2 \ell_0^{(d)}(w(\varphi);z(\varphi);\eta)
\end{aligned}
\end{equation}
and
\begin{equation}
\begin{aligned}
\label{eq_beta_fieldrenormalization}
&\beta_z^{(d)}(\varphi;\eta)\equiv\beta_z^{(d)}(w(\varphi),w'(\varphi);z(\varphi),z'(\varphi),z''(\varphi);\eta)\\&
= - \frac{4v_d}d w'(\varphi)^2m_{4,0}^{(d)}(w(\varphi);z(\varphi);\eta) - \\&
\frac {8v_d}d w'(\varphi)z'(\varphi)m_{4,0}^{(d+2)}(w(\varphi);z(\varphi);\eta)\\&
- \frac{4v_d}d z'(\varphi)^2 m_{4,0}^{(d+4)}(w(\varphi);z(\varphi);\eta) - 2v_d z''(\varphi) \times \\&
\ell_1^{(d)}(w(\varphi);z(\varphi);\eta)+  4 v_d u'''(\varphi) z'(\varphi) \ell_2^{(d)}(w(\varphi);z(\varphi);\eta) \\&
+ 2 v_d \frac{1+2d}d z'(\varphi)^2 \ell_2^{(d+2)}(w(\varphi);z(\varphi);\eta).
\end{aligned}
\end{equation}
The above beta functions can be more explicitly obtained by taking derivatives with respect to $w$ and $z$ of Eqs.~(\ref{eq_threshold_ell}) 
and (\ref{eq_threshold_m}). Note that $\beta_z^{(d)}$ does not depend on $w''$.

After inserting $w_k(\varphi)=w(\varphi) +k^\lambda \delta w(\varphi)$, $z_k(\varphi)=z(\varphi) +k^\lambda \delta z(\varphi)$, 
$\eta_k=\eta+k^\lambda \delta \eta$, where $w$, $z$, and $\eta$ are the fixed-point quantities, and linearizing the FRG flow equations, we 
obtain Eq.~(\ref{eq_linearized}) with
\begin{equation}
\begin{aligned}
&A_1(\varphi) = w(\varphi) +\frac 1{2}\varphi w'(\varphi)+ \partial_\eta \beta^{(d)}_{w}(\varphi;\eta),\\&
A_2(\varphi) = z(\varphi) +\frac 1{2}\varphi z'(\varphi)+ \partial_\eta \beta^{(d)}_{z}(\varphi;\eta),
\end{aligned}
\end{equation}
as well as
\begin{equation}
\begin{aligned}
&a_{11}(\varphi) = \partial_w \beta^{(d)}_{w}(w,w',w'';z,z',z''\eta),\\&
a_{12}(\varphi) = \partial_z \beta^{(d)}_{w}(w,w',w'';z,z',z''\eta),\\&
a_{21}(\varphi) = \partial_w \beta^{(d)}_{z}(w,w';z,z',z''\eta),\\&
a_{22}(\varphi) = \partial_z \beta^{(d)}_{z}(w,w';z,z',z''\eta),
\end{aligned}
\end{equation}
where $w$, $w'$, $w''$, $z$, $z'$, and $z''$ are formally considered as independent variables.

Similarly,
\begin{equation}
\begin{aligned}
&b_{11}(\varphi) = \partial_{w'} \beta^{(d)}_{w}(w,w',w'';z,z',z''\eta),\\&
b_{12}(\varphi) = \partial_{z'} \beta^{(d)}_{w}(w,w',w'';z,z',z''\eta),\\&
b_{21}(\varphi) = \partial_{w'} \beta^{(d)}_{z}(w,w';z,z',z''\eta),\\&
b_{22}(\varphi) = \partial_{z'} \beta^{(d)}_{z}(w,w';z,z',z''\eta),
\end{aligned}
\end{equation}
and 
\begin{equation}
\begin{aligned}
&c_{11}(\varphi) = \partial_{w''} \beta^{(d)}_{w}(w,w',w'';z,z',z''\eta),\\&
c_{12}(\varphi) = \partial_{z''} \beta^{(d)}_{w}(w,w',w'';z,z',z''\eta),\\&
c_{21}(\varphi) = \partial_{w''} \beta^{(d)}_{z}(w,w';z,z',z''\eta),\\&
c_{22}(\varphi) = \partial_{z''} \beta^{(d)}_{z}(w,w';z,z',z''\eta).
\end{aligned}
\end{equation}

More explicitly, the $c_{ij}$'s are given by
\begin{equation}
\begin{aligned}
&c_{11}(\varphi) =- 2v_d \ell_1^{(d)}(w(\varphi);z(\varphi);\eta),\\&
c_{12}(\varphi) = 2v_d \partial_z\ell_0^{(d)}(w;z;\eta)\vert_{w=w(\varphi), z=z(\varphi)},\\&
c_{21}(\varphi) = 0,\\&
c_{22}(\varphi) =  - 2v_d \ell_1^{(d)}(w(\varphi);z(\varphi);\eta),
\end{aligned}
\end{equation}
so that one checks that the matrix is invertible with determinant $[2v_d \ell_1^{(d)}(w(\varphi);z(\varphi);\eta)]^2>0$. The explicit 
expressions of the other functions are also trivially derived but are not worth displaying.

From the above equations, one can see that by construction, the $A_i$'s, $a_{ij}$'s and $c_{ij}$'s are even in $\varphi$ while the 
$b_{ij}$'s are odd.
\\

\section{Two zero eigenmodes in the LPA' approximation when $\tilde\epsilon\to 0$}
\label{app_eigenvalue-LPA'}

We revisit here the leading eigenvalues around the fixed point in the LPA' approximation when $\tilde\epsilon \to 0$. The equations to be 
considered are the flow equation for the square-mass function $w_k(\varphi)$ and for the running anomalous dimension $\eta_k$ 
which is defined by a prescription at the running minimum $\varphi_{{\rm min},k}$ of the potential,\cite{lnf23}
\begin{equation}
\label{eq_etak}
\eta_k= \frac{4v_d}d w'(\varphi_{{\rm min},k})^2m_{4,0}^{(d)}(w(\varphi_{{\rm min},k});\eta_k) 
\end{equation}
where $m_{4,0}^{(d)}(w;\eta)$ is given by Eq.~(\ref{eq_threshold_m}) with $n=4$ and $z=1$. We study the perturbation around the fixed 
point $w_k(\varphi)=w(\varphi)+k^\lambda \delta w(\varphi)$ and $\eta_k=\eta+k^\lambda \delta\eta$, and we do so through a singular 
perturbation treatment: see also [\onlinecite{lnf23}].

In the boundary layer there is a solution at leading order when $\tilde\epsilon \to 0$ with eigenvalue $\lambda=0$ and 
$\delta w(\varphi)\propto w'(\varphi)$. One can easily check that this corresponds also to $\delta\eta=0$.

The issue to be resolved is how can one extend this eigenmode to the whole range of variation of $\varphi$. We therefore consider 
the eigenvalue equation for $\delta w(\varphi)$ associated with $\lambda=0$ and $\delta\eta=0$ in the region where $\varphi={\rm O}(1)$ 
and formally setting $\tilde\epsilon=0$. The equation is then
\begin{equation}
\label{eq_eigenvalue_LPA'}
0= \delta w(\varphi)-\frac{2v_d}{d} \partial^2_\varphi[-\ell^{(d)}_1(\overline w(\varphi))\delta w(\varphi)] .
\end{equation}
where we have used that $\eta=2-d+{\rm O}(\tilde\epsilon)$ and that $\partial_w \ell^{(d)}_0(w)=-\ell^{(d)}_1(w)$. The corresponding 
equation for the fixed-point function is the same as Eq.~(\ref{eq:fp_w}) with $z=1$, and it has a periodic solution $\overline w(\varphi)$ 
with half-period $\varphi_*$ which becomes asymptotically close to $\varphi_{\rm min}$ and diverges as $\sqrt{\ln(1/\tilde\epsilon)}$ 
when $\tilde\epsilon\to 0$.

We look for a solution of Eq.~(\ref{eq_eigenvalue_LPA'}) of the form $\delta w(\varphi)=c(\varphi)\overline w'(\varphi)$. We then arrive 
at the equation
\begin{equation}
\label{eq:Fequation}
0=2\partial^2_\varphi \ell^{(d)}_0(\overline w(\varphi)) c'(\varphi)+\partial_\varphi \ell^{(d)}_0(\overline w(\varphi)) c''(\varphi),
\end{equation}
whose generic solution is
\begin{equation}
\label{eq:Fequation}
c'(\varphi)=\frac{A}{[\partial_\varphi \ell^{(d)}_0(\overline w(\varphi))]^2},
\end{equation}
with $A$ some constant. We immediately see that $A=0$ corresponding to $c(\varphi)=K$ is a solution and corresponds 
to an odd eigenvector $\delta w(\varphi)$ under the inversion of $\varphi$. This eigenvector can be matched with the solution in the 
boundary layer and it is associated with the eigenvalue $\lambda=-(d-2+\eta)/2\propto \tilde\epsilon \to 0$: see Sec.~\ref{sub_generalities}.
\\

When $A\neq 0$, $c'(\varphi)$ is an even function that diverges as $1/\varphi^2$ when $\varphi \to 0$. Furthermore it  varies slowly and is 
vanishingly small in the whole region $1\ll \varphi\ll \varphi_*$. One can therefore choose a primitive $c(\varphi)$ which is odd and behaves 
as $1/\varphi$ when $\varphi \to 0$,
\begin{equation}
\begin{aligned}
c(\varphi)\propto -\frac 1{\varphi}+
\int_0^\varphi d\varphi' \Big (\left [\frac{\ell^{(d)}_1(\overline w_0)\overline w''_0}{\ell^{(d)}_1(\overline w(\varphi'))\overline w'(\varphi')}\right ]^2 
-\frac 1{\varphi'^2}\Big )
\end{aligned}
\end{equation}
where the integral is convergent when $\varphi\to 0$ as $\overline w'(\varphi')=\overline w''_0 \varphi +{\rm O}(\varphi^2)$. This leads to 
the behavior described in Eq.~(\ref{eq:c_limits}). The resulting eigenfunction $f(\varphi)=c(\varphi)\overline w'(\varphi)$ is even and is 
well defined in $\varphi=0$ (the $1/\varphi$ of $c(\varphi)$ is then compensated by  $w'(\varphi)\propto \varphi$).
\\

\bibliographystyle{apsrev4-1}
\bibliography{bibli}

\end{document}